\documentclass[aps,pre,twocolumn,groupedaddress]{revtex4-1}

\usepackage{graphicx} 
\usepackage{float}
\usepackage[english]{babel}
\usepackage{mathtools}
\usepackage{bm}
\usepackage{amsmath}
\usepackage{amssymb}
\usepackage{color}
\usepackage[percent]{overpic}

\usepackage{epstopdf}%This line makes .pdf figures into .pdf - please comment out if not required.

\definecolor{cream}{RGB}{222,217,201}

\begin{document}

\title{\textbf{Binding of thermalized and active membrane
curvature-inducing proteins}} %Article title goes here instead of the text "This is the title"

\author{Quentin Goutaland}
\author{Fr\'ed\'eric van Wijland}
\author{Jean-Baptiste Fournier}
\affiliation{Laboratoire Mati\`ere et Syst\`emes Complexes (MSC), Universit\'e de Paris \& CNRS, 75013 Paris, France}
\author{Hiroshi Noguchi}
\affiliation{
Institute for Solid State Physics, University of Tokyo,
 Kashiwa, Chiba 277-8581, Japan}
\affiliation{Institut Lumi{\`e}re Mati{\`e}re, UMR 5306 CNRS, Universit{\'e} Lyon 1, F-69622 Villeurbanne, France}

\date{\today}

  \begin{abstract}
The phase behavior of the membrane induced by the binding of curvature-inducing proteins is studied by a combination of analytical and numerical approaches.
In thermal equilibrium under the detailed balance between binding and unbinding, 
the membrane exhibits three phases: an unbound uniform flat phase (U), a bound uniform flat phase (B), and a separated/corrugated phase (SC).
In the SC phase, the bound proteins form hexagonally-ordered bowl-shaped domains.
The transitions between the U and SC phases and between the B and SC phases are second order and first order, respectively.
At a small spontaneous curvature of the protein or high surface tension, the transition between  B and SC phases becomes continuous.
Moreover, a first-order transition between the U and B phases is found at zero spontaneous curvature driven by the Casimir-like interactions
between rigid proteins.
Furthermore, nonequilibrium dynamics is investigated by the addition of active binding and unbinding at a constant rate.
The active binding and unbinding processes alter the stability of the SC phase.
  \end{abstract}
   
\maketitle

%\preprint{}

%Title of paper

\section{Introduction}

In the biological realm, biomembranes can be found in a variety of shapes. These are regulated by a myriad of proteins~\cite{mcma05,shib09,baum11,mcma11,suet14,joha15}. Among these, some bind to the membrane and locally bend it. For instance, 
in endo/exocytosis, the formation of a spherical bud is directly scaffolded by binding clathrin and other proteins~\cite{mcma11,suet14,joha15}. The BAR superfamily of proteins is another example. These proteins bind to the membrane and bend it along the axis of their BAR domain thus leading to the formation of a cylindrical tube {\it in vivo} and {\it in vitro}~\cite{mcma05,suet14}.

In living cells, biomembranes evolve in nonequilibrium conditions, and the individual protein binding/unbinding is often activated by chemical reactions. For example, the clathrin coat of a vesicle is disassembled by ATP synthesis~\cite{mcma11}.
Dynamin forms a helical assembly on the membrane and induces fission by GTP binding and hydrolysis~\cite{schm11}.
BAR proteins contain binding sites or regulatory domains for GTPases or actin regulatory proteins~\cite{itoh05,aspe09}. Finally, the fluctuations of  membranes are often found to deviate from the equilibrium spectrum~\cite{pros96,turl16}.

In terms of collective behavior, spatiotemporal patterns in membranes are often observed in cell migration, spreading, growth, or division~\cite{yang18,dobe06,tani13,hoel16,kohy19}. Traveling waves of binding of F-BAR proteins and actin were experimentally observed~\cite{wu18}. 

In numerical simulations {in thermal} equilibrium, the protein binding to the membrane has been studied from molecular resolutions~\cite{bloo06,yu13,mahm19} 
to large-scale thin-surface membrane models~\cite{nogu17,hu11,sree18,gozd12,tozz19,rama18,nogu16a,Krishnan19,nogu19a}. When proteins generate an isotropic spontaneous curvature, bound sites are numerically found to assemble 
into circular domains or spherical buds~\cite{hu11,sree18,gozd12,tozz19,nogu16a}.
On the other hand, numerically still, bound sites with anisotropic spontaneous curvature have been found to induce membrane tubulation~\cite{rama18,nogu16a,nogu19a}.

In theoretical studies, the binding of proteins and their assembly on membranes have been explored in equilibrium for fully flat membranes~\cite{Chatelier96,Minton01,Zhdanov10} and in the more complicated case of curved membranes where the effects of a fixed curvature sensing~\cite{Singh12,Wasnik15,Krishnan19}  are present. But the phase diagram for a membrane which can freely curve in equilibrium and to which curvature-inducing proteins can bind and interact via curvature-mediated interactions~\cite{Goulian93EPL,Weikl98PRE,domm99,domm02,nogu17} deserves to be thoroughly investigated.

\begin{figure}\centering
\includegraphics[width=8cm]{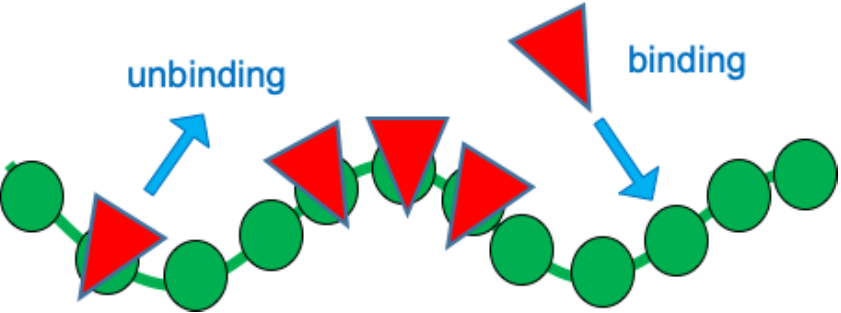}
\caption{
Binding and unbinding of molecules with a finite spontaneous curvature to the membrane.
}
\label{fig:bind}
\end{figure}

It is important to determine how the biologically relevant case of \textit{active} binding/unbinding, and the related nonequilibrium collective phenomena, affect the phase diagram of the membrane/proteins system and modifies the membrane shapes and domain structures. 
In this study, we thus examine the binding/unbinding of {proteins or other macromolecules} onto a deformable membrane, whether in or out of thermal equilibrium, using theory and simulations. We assume that the binding of {the molecule} locally changes the bending rigidity and induces a spontaneous curvature $C_0$ of the membrane, as shown in Fig.~\ref{fig:bind}. This spontaneous curvature is assumed to be isotropic, i.e., the bound {membrane} has no preferred bending orientation.  Theoretically, we resort to the Gaussian approximation of the Canham--Helfrich membrane curvature energy~\cite{canh70,helf73}, using a mixture of two different forms in order to take into account the mixture of bound and unbound {states}.  We incorporate in our description the {mixing} entropy and a chemical potential describing a binding/unbinding exchange with the bulk~\cite{baum11,Krishnan19}. The {membrane} dynamics is of diffusive nature. To describe them out-of-equilibrium, we use a noiseless Dean-Kawasaki equation~\cite{dean1996,KAWASAKI94}, together with  equilibrium and possibly active binding/unbinding processes. For simulations, we employ a meshless membrane model~\cite{nogu09,shib11,nogu14}, 
in which membrane particles self-assemble into a membrane. This model is adapted to the study of large-scale membrane dynamics (albeit solvent free) and is tunable in terms of the membrane's elastic properties.

In Section~\ref{sec:theory}, the theoretical analysis of the protein binding/unbinding is presented. It leads to a phase diagram, in or out of equilibrium, for the membrane phases in terms of $C_0$ and of the chemical potential.
In Section~\ref{sec:method}, the simulation model and method are described. In Sections~\ref{sec:thermal} and \ref{sec:active}, the simulation results of the thermal binding/unbinding process 
without and with active unbinding are presented, respectively. Comparison with the theoretical analysis shows good agreement. An outlook is presented in Section \ref{sec:sum}.

\section{Theory}\label{sec:theory}

We consider an incompressible membrane of fixed surface area $A$, which contains a surface density $\rho(\bm x)$ of bound proteins that are exchanged with a reservoir of chemical potential $\mu$. The membrane is assumed to be subjected to an external lateral tension $\gamma$ conjugate to the projected area $A_p=L_p^2$ (area of the membrane average plane); for the sake of simplicity, the corresponding contribution will be introduced at a later stage. We assume an ideal mixture between the protein-coated membrane and the bare membrane, so that the free energy of the system takes the form~\cite{Krishnan19}
\textcolor{black}{
\begin{align}
\label{eq:FFreeEnergy}
\mathcal{F}=\int_A\! &dS\,
\Big\{\!
\left(1-\rho a^2\right)
\left[
\frac{\kappa_u}2\left(c_1+c_2\right)^2+\bar\kappa_u c_1c_2
\right]
\nonumber\\
&+\rho a^2\left[
\frac{\kappa_b}2\left(c_1+c_2-C_0\right)^2
+\bar\kappa_bc_1c_2
\right]-\mu\rho
\nonumber\\
&+T\left[\rho\ln(\rho a^2)+\left( a^{-2}-\rho\right)\ln\left(1-\rho a^2\right)\right]
\!\Big\}.
\end{align}
}
The first term, proportional to $1-\rho a^2$, where $a^2$ is the surface area covered by a bound protein, is the standard Canham-Helfrich Hamiltonian describing the curvature energy of the unbound membrane fraction~\cite{Helfrich73}, with a bending rigidity $\kappa_u$ and a Gaussian modulus $\bar\kappa_u$. Like $\rho(\bm x)$, the membrane principal curvatures $c_1(\bm x)$ and $c_2(\bm x)$ are space-dependent. Since the stability of the bare flat membrane requires $-2\kappa_u<\bar\kappa_u<0$~\cite{Safran_book}, 
\textcolor{black}{
we shall assume $\bar\kappa_u=-\kappa_u$, as this value will also matches the numerical simulations.
}

The second term, proportional to $\rho$, is the standard Helfrich Hamiltonian for the bound membrane fraction, with a spontaneous curvature $C_0$.
\textcolor{black}{
We shall also choose $\bar\kappa_b=-\kappa_b$. In this case the term in brackets reduces to $\frac12\kappa_b(c_1-C_0)^2+\frac12\kappa_b(c_2-C_0)^2$ up to a constant  that we may discard as it simply renormalizes the chemical potential $\mu$. The proteins thus locally promote an isotropic curvature  of magnitude $C_0$, like conically-shaped inclusions would do. Although curvature-inducing proteins presumably stiffen the bound membrane (otherwise they would fail to impose a local prescribed curvature), 
their bending strength is protein-dependent. For the sake of simplicity, we shall assume $\kappa_b\sim 10\kappa_u$ in this study. }

The third term describes the equilibrium exchange of the proteins with the solvent, i.e., the binding/unbinding process, through a chemical potential $\mu$. Note that the binding energy has been absorbed in the definition of the chemical potential. The last term, with $T$ the temperature in energy units, describes the entropy of mixing of proteins.

For small deformations relative to the flat state, the shape of the membrane can be described in the Monge gauge by the height function $z=h(\bm r)$, where $\bm r$ covers a two-dimensional (2D) plane. To second order in the deformation $h$, we have then $c_1^2+c_2^2\simeq(\partial_i\partial_jh)^2$, $(c_1-C_0)^2+(c_2-C_0)^2\simeq(\partial_i\partial_jh-C_0\delta_{ij})^2$ and $dS\simeq[1+\frac12(\bm\nabla h)^2]d^2r$, where we used Einstein's summation convention, which will be implicit throughout. Thus, to second order, the free energy becomes $\mathcal{F}\simeq \tilde{\mathcal{F}}$, with
\begin{align}
\label{eq:Fprime}
\tilde{\mathcal{F}}=\int_{A_p}\!\!&d^2r\,
\bigg\{
\frac{(1-\rho a^2)\kappa_u+\rho a^2\kappa_b}2(\partial_i\partial_jh)^2
-\rho a^2\kappa_bC_0\bm\nabla^2h
\nonumber\\
&+\left(1+\frac12(\bm\nabla h)^2\right)
\bigg[\rho a^2\kappa_bC_0^2
+T\rho\ln(\rho a^2)
\nonumber\\
&+T\left( a^{-2}-\rho\right)\ln\left(1-\rho a^2\right)-\mu\rho\bigg]\bigg\},
\end{align}
where $A_p$ is the projected area of the membrane.
Note that $0<\rho a^2<1$ is not assumed to be small.

In this section, we will now work in dimensionless units by setting $ a=T=1$. In other words, we take $T$ as the unit of energy and $ a$ as the unit of length. In addition, since we are interested in a system with a large projected membrane area $A_p$ (and in accordance with the numerical simulations below), we will assume periodic boundary conditions.

\subsection{Linear stability analysis at equilibrium}\label{sec:lsa}

Due to the curvature promoted by the bound proteins, we expect the flat membrane to develop spatial undulations.
The only term that may destabilize the flat membrane, however, is the second term of eqn~\eqref{eq:Fprime}; but if $\rho$ is uniform, it is a boundary term with no effect under periodic boundary conditions.
\textcolor{black}{In other words, the energy gain in the favorably curved parts would be compensated by the energy loss in the unfavorably curved  parts.
}
We thus expect that at both low and high protein densities, the membrane will remain flat since the entropy of mixing will promote uniform density. At intermediate densities, however, lateral phase separation accompanied by spatial modulations of the membrane will be possible.
\textcolor{black}{
We therefore foresee three phases: an unbound uniform flat phase (U), i.e., a flat membrane with a low density of bound proteins, a bound uniform flat phase (B), i.e., a flat membrane with a high density of bound proteins, and a separated/corrugated phase (SC), i.e., a corrugated membrane with regions of different protein densities and curvatures. Since the U and B phases have the same symmetries, we expect either a first-order transition between them, or a transition through an intermediate phase, or a continuous transformation similar to the gas--liquid transformation above the critical point.
}

Let us first examine the situation where the membrane is flat. For $h=0$, the energy $ \tilde{\mathcal{F}}$ becomes 
\begin{align}
  \tilde{\mathcal{F}}_0=L_0^2\left[\rho\kappa_bC_0^2+\rho\ln\rho+(1-\rho)\ln(1-\rho)-\mu\rho\right] \label{eq:freeF},
\end{align}
where $L_0^2=A$ is the membrane area. Minimizing it with respect to $\rho$ gives the equilibrium density~\cite{Krishnan19}
\begin{align}
\rho_0=\frac1{1+\mathrm e^{\kappa_bC_0^2-\mu}},
\label{eq_density}
\end{align}
for which $ \tilde{\mathcal{F}}_0=L_0^2\ln(1-\rho_0)$.

We  now perform a linear stability analysis of this solution for a square membrane under an external lateral tension $\gamma$.
Let us consider a small perturbation $h=h_1(\bm r)$ and $\rho=\rho_0+\rho_1(\bm r)$ of the previous solution. Calling  $L_p=L_0+L_1$ the linear size of the perturbed membrane, the area constraint reads $L_p^2+\int_0^{L_p}d^2r\frac12(\bm \nabla h)^2= L_0^2$, yielding to second order in the perturbation 
\begin{align}
\label{eq:L1}
L_1\simeq-\frac1{2L_0}\int_0^{L_0}\!\mathrm d^2r\,
 \frac12(\bm\nabla h_1)^2.
\end{align}
Taking into account that $\kappa_bC_0^2-\mu=\ln[(1-\rho_0)/\rho_0]$, the energy becomes at second order in the perturbation
\begin{align}
\label{eq:Fprim2}
 \tilde{\mathcal{F}}\simeq\int_0^{L_p}\!&d^2r\,
\Big[
\frac{\kappa_\mathrm{eff}}2(\bm\nabla^2h_1)^2
-\rho_1\kappa_bC_0\bm\nabla^2h_1
\nonumber\\&
+\Big(1+\frac12(\bm\nabla h_1)^2\Big)\ln(1-\rho_0)
+\frac{\rho_1^2}{2\rho_0(1-\rho_0)}
\Big],
\end{align}
where $\kappa_\mathrm{eff}=(1-\rho_0)\kappa_u+\rho_0\kappa_b$. Note that we have discarded a term $\propto\!\bm\nabla^2h_1$ that vanishes under periodic boundary conditions, and replaced $(\partial_i\partial_jh)^2$ by $(\bm\nabla^2h)^2$, since the difference vanishes under periodic boundary conditions. 

\begin{figure}
\centerline{\includegraphics[width=.8\columnwidth]{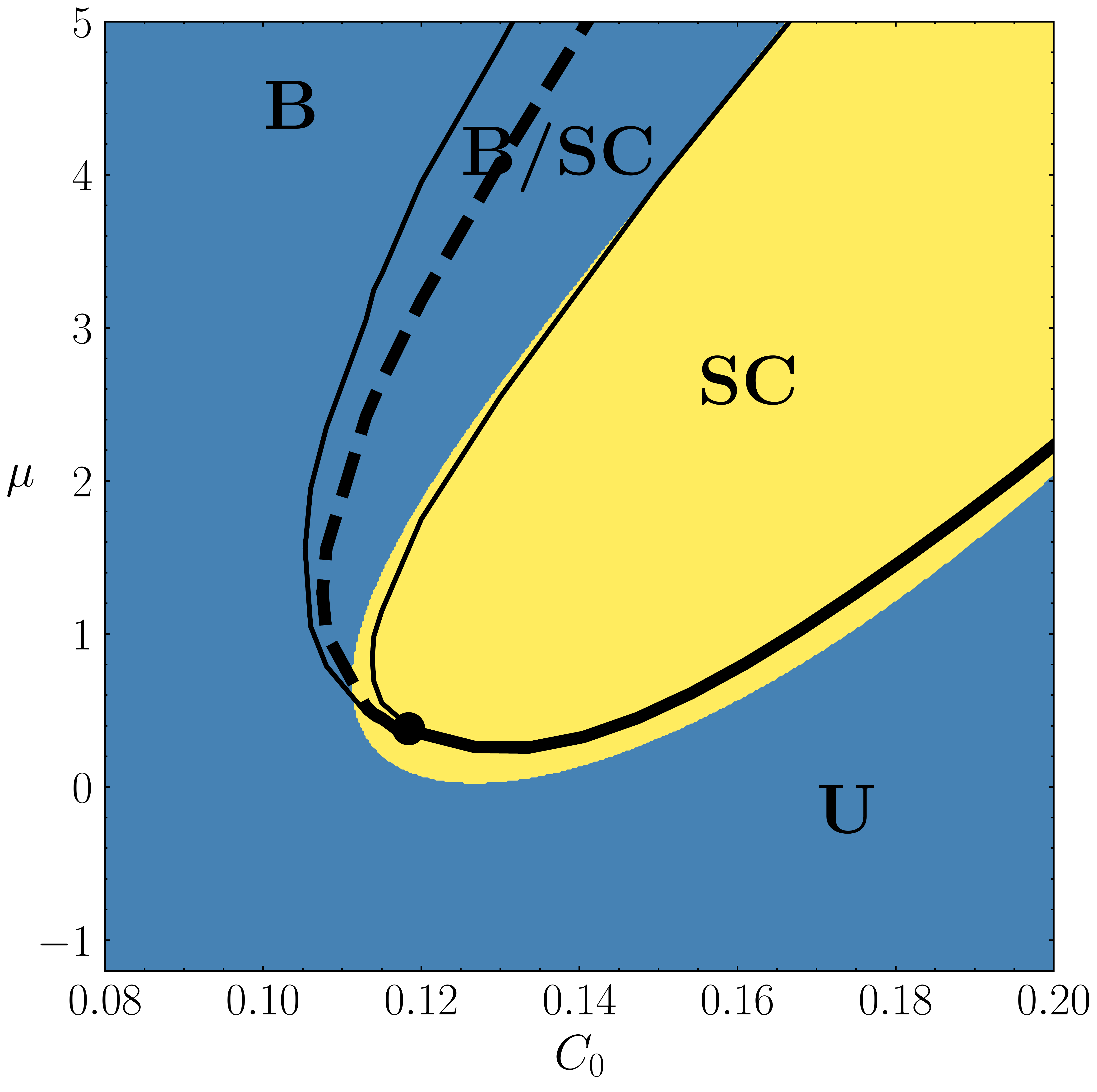}}
\caption{Equilibrium phase diagram. In the {blue region,} the flat membrane is stable against small perturbations, while in the {yellow region,} it is unstable. The black lines show the phase diagram obtained from the bumpy 1D shapes studied in the nonlinear analysis. U: unbound flat phase (low protein density), B: bound flat phase (high protein density), SC: separated-corrugated phase where the membrane exhibits curved domains with a separation between protein-dense and protein-poor regions. The thick solid line indicates a second-order phase transition. The thick dashed line corresponds to a first-order phase transition, with a coexistence region delimited by the two thin solid lines. The black dot indicates a tricritical point. Parameters are $\kappa_u=16$, $\kappa_b=144$ and $\gamma=0.5$.}
\label{fig:eq_ph_diag}
\end{figure}

Now, to second order in the perturbation, we have
\begin{align}
\int_0^{L_p}\!\!d^2r
\ln(1-\rho_0)\simeq(L_0^2+2L_0L_1)\ln(1-\rho_0),
\end{align}
therefore this term gives $ \tilde{\mathcal{F}}_0=L_0^2\ln(1-\rho_0)$ plus a contribution that cancels the term proportional to $(\bm\nabla h_1)^2$ in eqn~\eqref{eq:Fprim2}, because of eqn~\eqref{eq:L1}. Adding the energy associated with the external tension, the total energy becomes $\mathcal F^\star= \tilde{\mathcal{F}}-\gamma L_p^2$, which reads, up to a constant and at second order in the perturbation,
\begin{align} 
\label{eq:Fstar}
\mathcal F^\star\simeq
\int_0^{L_0}&d^2r
\Big[
\frac{\kappa_\mathrm{eff}}2(\bm\nabla^2h_1)^2
-\rho_1\kappa_bC_0\bm\nabla^2h_1
\nonumber\\&
+\frac12\frac{\rho_1^2}{\rho_0(1-\rho_0)}
+\frac\gamma2(\bm\nabla h_1)^2  
\Big].
\end{align}
Calling $h_{1,\bm q}$ and $\rho_{1,\bm q}$ the Fourier transforms of the perturbation fields, we obtain
\begin{align}
\label{eq:FFourier}
\mathcal F^\star\simeq
\frac12L_0^2
\sum_{\bm q}
\begin{pmatrix}
h_{1,\bm q} \\                                          
\rho_{1,\bm q}                                         \end{pmatrix}^t\!
\mathsf M_\mathrm{eq}
\begin{pmatrix}
h_{1,-\bm q} \\                                          
\rho_{1,-\bm q}                                         \end{pmatrix},
\end{align}
with 
\begin{align}
\mathsf M_\mathrm{eq}=
\begin{pmatrix}
\kappa_\mathrm{eff}q^4+\gamma q^2 & \kappa_bC_0q^2\\
\kappa_bC_0 q^2 & \frac1{\rho_0(1-\rho_0)}
\end{pmatrix}.
\label{Meq}
\end{align}

The flat membrane is unstable if $\mathsf M_\mathrm{eq}$ has negative eigenvalues. Since $\mathop{\mathrm{tr}}(\mathsf M_\mathrm{eq})>0$, the corresponding condition is $\det(\mathsf M_\mathrm{eq})<0$, which reads
\begin{align}
\left[\kappa_\mathrm{eff}-\kappa_b^2C_0^2\rho_0(1-\rho_0)\right]q^2+\gamma <0.
\label{detM2}
\end{align}
{For tensionless membrane ($\gamma=0$), all $q$ modes are therefore destabilized when the quantity $\delta=\kappa_\mathrm{eff}-\kappa_b^2C_0^2\rho_0(1-\rho_0)$ in the square brackets above is negative.
For $\gamma>0$, the unstable modes} are in the interval $q\in [q_{\rm min},q_{\rm max}]$, with $q_{\rm max}=+\infty$ and $q_{\rm min}=\sqrt{{\gamma}/{(-\delta)}}$. Now, for membranes with proteins, our length scale $a$ also corresponds to the smallest wavelength accessible to membrane fluctuations, which sets an upper cutoff in Fourier space of order $1/a$, or, in dimensionless units, of order $1$. Thus, if $q_{\rm min}$ is larger than $1$ there is no physical range of $q$ that can be excited by the instability, even if $\delta<0$. Hence, we expect that when $\delta<-\gamma$ there will be a separated and modulated phase (SC) if the condition $\delta<0$ is met. A necessary condition for $\delta<0$ is given by $C_0>C_\mathrm{th}$, with
\begin{align}
C_\mathrm{th}=\frac{\sqrt{\kappa_u}+\sqrt{\kappa_b}}{\kappa_b}.
\end{align}
{The instability condition  $\delta<-\gamma$} is actually fulfilled in the range of $\mu$ shown in the yellow region of  Fig.~\ref{fig:eq_ph_diag}. {This unstable region is slightly shrunk for small values of $\gamma$.}
As expected, spatial undulations occur when the membrane is neither too poor nor too rich in proteins, i.e., at intermediate values of $\mu$ where the entropy of the mixture allows phase separation, as evidenced by the presence of the $\rho_0(1-\rho_0)$ factor in the instability condition.

\begin{figure}
\centerline{\includegraphics[width=.8\columnwidth]{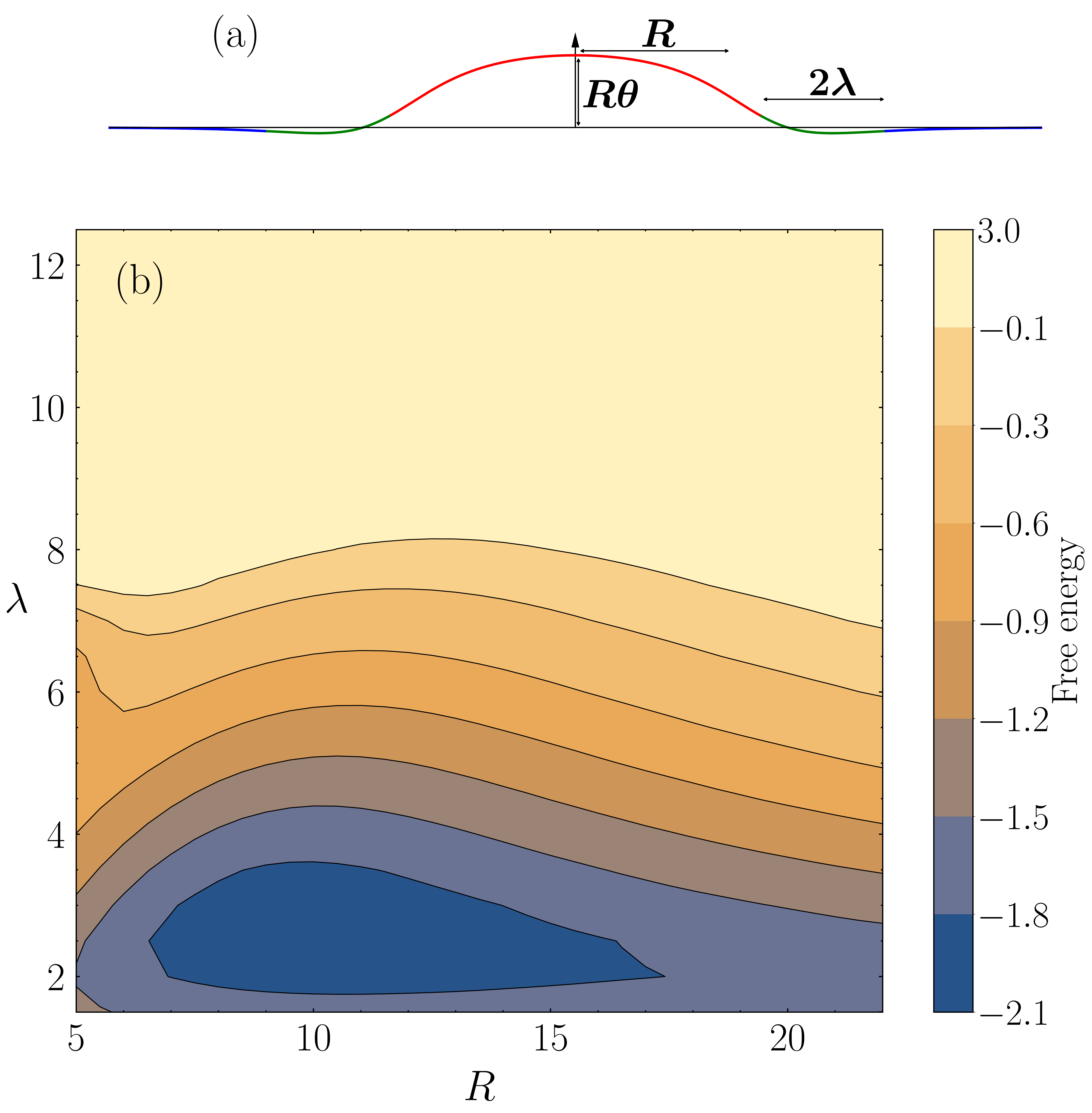}}
\caption{(a) Typical membrane shape with a bump for use in the nonlinear analysis (cross section). (b) Free energy $f$ of the bump as a function of $R$ and $\lambda$ after numerical minimization with respect to $\theta$. The blue regions of low energy correspond to $\lambda\ll R$. Parameters are $\kappa_u=16$, $\kappa_b=144$, $\gamma=0.5$, $C_0=0.15$ and $\mu=2$ (deep in the instability region).
}
\label{fig:hump}
\end{figure}

\subsection{Nonlinear analysis in equilibrium}\label{sec:naeq}

The linear stability analysis is of course unable to predict the corrugation pattern selected by the system in the nonlinear regime. Instead of solving the corresponding nonlinear PDE's (with the covariant Helfrich contribution to the energy) we follow an alternative route. We postulate that the system will adopt a corrugated phase, which we parametrize with a small number of parameters. To simplify, we assume in addition that the selected patterns are translationally invariant along one space direction. 

To study the phase diagram of the system beyond the linear stability analysis, we thus consider a family of corrugated shapes of arbitrary amplitude (Fig.~\ref{fig:hump}a). In the SC phase, we expect the system to develop periodic structures with large regions of curvature favorable to inclusions, surrounded by narrow regions of opposite curvature. Proteins will naturally accumulate in the favorable regions and deplete in the unfavorable regions. We thus consider the following three-parameter family of smooth shapes:
\begin{align}
h(x)=R\theta \,
\mathop{\Gamma}\left(\frac xR\right)
\frac{\tanh\left(\frac{x+R}{\lambda}\right)-\tanh\left(\frac{x-R}{\lambda}\right)}{2\tanh\left(\frac{2 R}{\lambda}\right)^2},
\end{align}
with $\Gamma(x)$ the circular arc of equation $\sqrt{R^2-x^2}$. This function, parametrized by ($R,\lambda, \theta$), describes a membrane deformation having a central circular bump of width $2R$ and amplitude $R\theta$ surrounded by side channels of width $2\lambda$ in which the curvature changes sign and relaxes (Fig.~\ref{fig:hump}a). This deformation can be repeated in space in order to produce a corrugation.

We then seek to determine the equilibrium state of the system. With $c_1=c$ and $c_2=0$,
the free energy~\eqref{eq:FFreeEnergy} per unit length, supplemented by the contribution of the external tension, takes the exact nonlinear form:
\begin{align}
f=
\int_{-\frac{L_p}2}^{{\frac{L_p}2}} &\!\sqrt{1+h'^2}\,
\Big[
(1-\rho)\frac{\kappa_u}2c^2
+\rho\frac{\kappa_b}2[(c-C_0)^2+C_0^2]
\nonumber\\&+
\rho\ln\rho+(1-\rho)\ln(1-\rho)-\mu\rho \Big]dx-\gamma L_p,
\end{align}
with $c=h''/(1+h'^2)^{3/2}$,
where $L=\int_{-L_p/2}^{L_p/2}dx\,\sqrt{1+h'^2}$,
is the fixed total length perpendicular to the translationally invariant direction and $L_p$ is its variable projected length determined consistently. Constructed this way, $f$ is a function of $R$, $\lambda$, $\theta$ and a functional of $\rho(x)$. Minimizing with respect to $\rho(x)$ gives $\rho^\star(x)=[1+\exp(-\frac12\kappa_uc(x)^2+\frac12\kappa_b[(c(x)-C_0)^2+C_0^2]-\mu)]^{-1}$, and the free energy per unit length reduces to
\begin{align}
f&=\int_{-\frac{Lp}2}^{{\frac{Lp}2}}\sqrt{1+h'^2}\left[\frac\kappa2c^2+\ln(1-\rho^\star) \right]dx-\gamma L_p.
\end{align}

Let us place ourselves in the region of instability of the linear stability analysis (Fig.~\ref{fig:eq_ph_diag}), and numerically examine  when the bumps described by $f(R,\lambda,\theta)$ are stable with respect to the flat state. Scanning the ($R$,$\lambda$) plane and  minimizing numerically the energy with respect to $\theta$ (Fig.~\ref{fig:hump}b), we find that the stable bumps appear in a large $R$ range but have a small, well-defined $\lambda$ value. We therefore expect the SC phase to consist of large protein-filled bumps surrounded by narrow protein-depleted oppositely curved channels.

\begin{figure}[h]
\centerline{\includegraphics[width=.8\columnwidth]{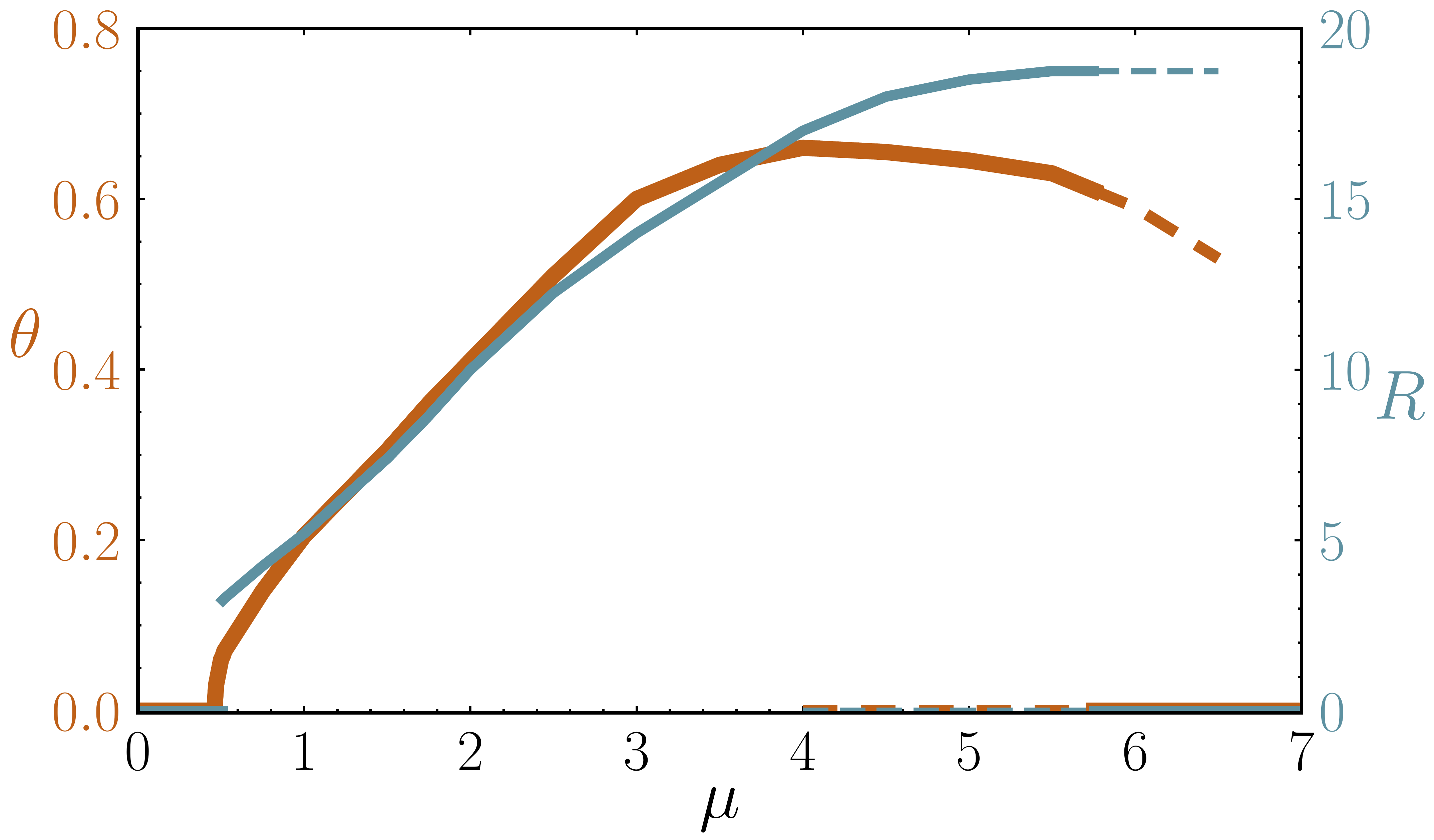}}
\caption{Equilibrium amplitude $\theta$ (order-parameter) and width $R$ of the 1D bumps used in the nonlinear analysis as a function of $\mu$. The solid lines correspond to stable states and the dashed lines to metastable states. Parameters are the same as in Fig.~\ref{fig:eq_ph_diag} and $C_0=0.15$. }
\label{fig:orderparams}
\end{figure}

Since the system prefers narrow side channels, we set a small $\lambda\simeq2.5$ (see Fig.~\ref{fig:hump}) and we keep $R$ and $\theta$ as parameters. We find that the one-dimensional  bumps shapes, corresponding to the separated/corrugated (SC) phase, are stable with respect to the flat state within the region delimited by the thick solid and thick dashed lines in Fig.~\ref{fig:eq_ph_diag}. Starting at small chemical potentials and increasing $\mu$, the transition from the flat unbound phase (U) to the separated/corrugated phase (SC) is of second order to the right of a tricritical point and of first order to its left. Increasing $\mu$, there is, as expected, a re-entrant first-order transition toward a flat state, the flat bound phase (B). The equilibrium values of $R$ and $\theta$ are shown in Fig.~\ref{fig:orderparams}.

\subsection{Linear stability analysis in the presence of active binding/unbinding}\label{sec:lsa}

In nonequilibrium, it is necessary to specify the dynamics of the system to study its behavior.
We model the binding-unbinding mechanism by a Poisson process with rates $\eta_1$ and $\eta_2$~:
\begin{equation}\label{eq:cartoon-activity}
\text{bound}\xrightleftharpoons[\eta_1]{\eta_2}\text{unbound}
\end{equation}

\textcolor{black}{
Neglecting the fluctuations caused by the binding/unbinding active processes and by the thermal exchanges with the thermostat, we consider the following 
noiseless dynamical 
equations for the density and height fields:
}
\begin{align}
\label{eq:eqdotrho}
\dot{\rho}(\bm r)&=\bm\nabla\cdot\left(m\rho(1-\rho a^2)\bm\nabla\frac{\delta \hat{\mathcal{F}}}{\delta\rho}\right)+ \alpha_1( a^{-2}-\rho) 
\nonumber\\
& -\alpha_2\rho
+\eta_1( a^{-2}-\rho)-\eta_2\rho, \\
 \dot h(\bm r)&=-\Lambda\frac{\delta\hat{\mathcal F}}{\delta h},
 \label{eq:eqdoth}
\end{align}
where $\hat{\mathcal{F}} = \tilde{\mathcal{F}}-\gamma L_p^2$.
The first term in eqn~\eqref{eq:eqdotrho}, proportional to the  mobility $m$ of the particles, stems from the conservation of their number in the absence of binding/unbinding processes. It is the divergence of the particle current, in which the $\rho(1-\rho a^2)$ factor accounts for the vanishing of the current both for $\rho=0$ (empty state) and $\rho=a^{-2}$ (filled state). Note that for small $\rho$ this part of the  equation reduces to the noiseless Dean-Kawasaki equation~\cite{Dean96JP,Kawasaki94PA}. The next two terms describe the equilibrium binding and unbinding of the proteins, respectively.
Note that there is some freedom in choosing the thermal binding-unbinding rates, as long as the detailed balance condition is fulfilled i.e., $\alpha_1/\alpha_2=e^{-\beta\Delta\mathcal{H}(\bm r)}$. One possibility, which we adopt, is to resort to the Glauber rates defined by $\alpha_1=\alpha e^{\beta\Delta\mathcal{H}(\bm r)}/(1+e^{\beta\Delta\mathcal{H}(\bm r)})$ and $\alpha_2=\alpha e^{-\beta\Delta\mathcal{H}(\bm r)}/(1+e^{-\beta\Delta\mathcal{H}(\bm r)})$ where $\Delta\mathcal{H}(\bm r)=\frac12\kappa_b(\partial_i\partial_jh-C_0\delta_{ij})^2-\frac12\kappa_u(\partial_i\partial_jh)^2-\mu$ is the energy variation upon binding of a protein. 
While also being associated to local configuration changes, they further share with the Metropolis rates used in the simulations (that we will present in Section~\ref{sec:method}) the property that they remain bounded, regardless of the energy change involved. Other choices are of course possible. While these various choices leave the equilibrium phase diagram intact, the resulting nonequilibrium steady state in the presence of active processes will depend on the specific choice that is made. However, we have checked that alternative choices \textcolor{black}{involving only local moves}, {\it e.g.} $\alpha_1 = \alpha e^{-\frac12\beta\Delta\mathcal H(\bm r)}$ and $\alpha_2 = \alpha e^{\frac12\beta\Delta\mathcal H(\bm r)}$ or $\alpha_1 = \alpha e^{-\beta\Delta\mathcal H(\bm r)}$ and $\alpha_2 = \alpha$, although they affect the specific location of phase and local stability boundaries, do not alter our physical conclusions for physically relevant values of $C_0$, $\mu$ and $\eta_1$ for which the effects of the active binding process do not dominate the ones of the thermal processes.  \textcolor{black}{Note that with our choice of rates that saturate when the energy difference increases by a large amount across a configuration change, we mimic the chemical/physical reality according to which diffusion constants are bounded}. 

The last two terms in eqn~\eqref{eq:eqdotrho} describe the active binding ($\eta_1$) and active unbinding ($\eta_2$) processes. Since  the rates $\eta_1$ and $\eta_2$ are constant, these terms violate detailed balance and constitute the source of nonequilibrium.  The second equation, which describes the dynamics of the membrane shape $h$, assumes a simple local dissipative dynamics with mobility 
$\Lambda$. Hydrodynamic interactions are thus neglected. We now switch to dimensionless units by setting $a=T=m=1$. In other words, we take $a^2/(\tilde\mu T)$ as the unit of time.

Let us first investigate the steady state for the flat uniform state. The dynamical equation for $\rho$ becomes
\begin{align}
\dot\rho=&
\alpha(1-\rho){\frac{e^{-(\kappa_bC_0^2-\mu)}}{1+e^{-(\kappa_bC_0^2-\mu)}}}
-\alpha\rho {\frac{e^{\kappa_bC_0^2-\mu}}{1+e^{\kappa_bC_0^2-\mu}}}
\nonumber\\
&+\eta_1(1-\rho)-\eta_2\rho,
\end{align}
and the steady-state solution is therefore
\begin{align}
 \bar\rho_0=\frac{\alpha\rho_0+\eta_1}{\alpha+\eta_1+\eta_2},
\end{align}
with $\rho_0$ given in eqn~\eqref{eq_density}. Note that $\bar\rho_0$ reduces as expected to $\rho_0$ in the absence of activity.

To perform a linear stability analysis in this nonequilibrium situation, we consider a small perturbation $h=h_1(\bm r)$, $\rho=\bar\rho_0+\rho_1(\bm r)$ and $L=L_0+L_1$, where $L_1$ is given by eqn~\eqref{eq:L1} in order to conserve the membrane area. To second order in the perturbation, we find that $\hat{\mathcal F}$ takes the same form as eqn~\eqref{eq:Fstar} except for an additional term $\rho_1\ln\{[\bar\rho_0(1-\rho_0)]/[\rho_0(1-\bar\rho_0)]\}$ in the integrand.
At first order in the perturbation, the dynamical equations~\eqref{eq:eqdotrho} and \eqref{eq:eqdoth} take then the form
\begin{align}
\begin{pmatrix}
\dot h_{1,\bm q} \\                                          
\dot\rho_{1,\bm q}                                         \end{pmatrix}
=- \mathsf M
\begin{pmatrix}
h_{1,\bm q} \\                                          
\rho_{1,\bm q}                                         \end{pmatrix},
\end{align}
with
\begin{align}
M_{11} &= \Lambda\left(\bar\kappa_\mathrm{eff}q^4+\gamma q^2\right),
\\
M_{12} &= \Lambda \kappa_bC_0q^2,
\\ 
M_{21} &= 
\kappa_bC_0q^2\left(
\bar sq^2
+ \alpha s\right),
\\
 M_{22} &= 
q^2+\eta_1+\eta_2+ \alpha,
\end{align}
where $\bar\kappa_\mathrm{eff}=(1-\bar\rho_0)\kappa_u+\bar\rho_0\kappa_b$, $s=\rho_0(1-\rho_0)$ and $\bar s=\bar\rho_0(1-\bar\rho_0)$.

The flat state is unstable when $\mathsf M$ has negative eigenvalues. Since $\mathrm{tr}(\mathsf M)$ is positive, this is achieved when $\det(\mathsf M)<0$. Let us first discuss the equilibrium case again. For $\eta_1=\eta_2=0$ and $\rho_0=\bar\rho_0$, we get $\mathsf M=\mathsf D\,\mathsf M_\mathrm{eq}$, with $\mathsf D=\mathop{\mathrm{diag}}(\Lambda,{s(q^2+\alpha)})$ and $M_\mathrm{eq}$ given by eqn~\eqref{Meq}. Since $\mathsf D$ is diagonal with strictly positive eigenvalues, the instability occurs when $\det(\mathsf{M}_\mathrm{eq})<0$ in agreement with the equilibrium condition of Section~\ref{sec:lsa}.

\begin{figure}
\centerline{\includegraphics[width=\columnwidth]{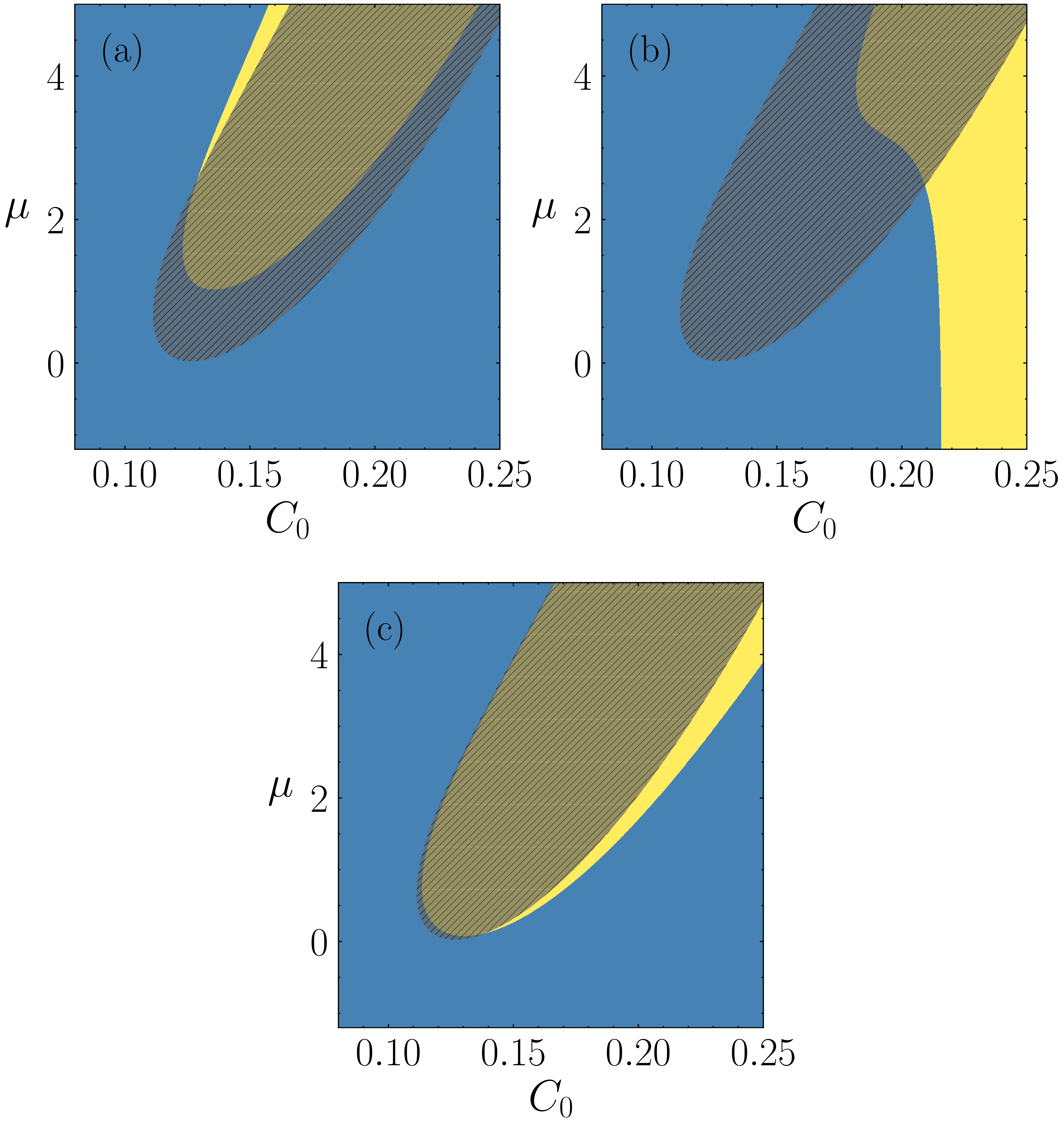}}
\caption{Linear stability analysis in the presence of active binding and active unbinding.  In the blue region, the flat membrane is stable against small perturbations, while in the yellow region, it is unstable. The hatched domain refers to the linearly unstable region in the equilibrium phase diagram of Fig.~\ref{fig:eq_ph_diag}. (a) Active unbinding for $\eta_1=0$ and $\eta_2=1$. (b) Active binding for $\eta_1=1$ and $\eta_2=0$. (c) Active binding for $\eta_1=0.02$, $\eta_2=0$. At small values of $\eta_1$ and $C_0$, there is no qualitative difference with (a) in terms of the shape of the instability region.
Parameters are $\kappa_u=16$ and $\kappa_b=144$, $\gamma=0.5$ and $\alpha=1$. Note that these phase diagrams are independent of $\Lambda$.}
\label{fig:activeLSA}
\end{figure}

In the general nonequilibrium case, the condition $\det(\mathsf M)<0$ yields an instability region that is shifted relative to the equilibrium case (Fig.~\ref{fig:activeLSA}). Both in the active binding and unbinding cases, the instability is shifted towards larger values of $C_0$, whereas it is shifted towards larger values of $\mu$ in the unbinding case and towards lower values of $\mu$ in the  binding case.\\

While our linear stability analysis predicts that the homogeneous phase is destabilized for active binding when $C_0$ exceeds a ($\eta_1$-dependent) threshold value (see Fig.~\ref{fig:activeLSA}(b)), we also note that the instability region has a similar shape for active binding and for active unbinding at moderate $C_0$ values, as seen in Fig.~\ref{fig:activeLSA}(c). However, we believe that the instability driven by a nonzero $\eta_1$ at large $C_0$ may not yield a simple SC phase, as more drastic nonlinear phenomena could take over in this regime. And as a matter of fact, in the simulations presented later in Section~\ref{sec:method}, right below eqn~\eqref{eq:Metro}, we have not considered the possibility of active binding, as the destabilization we have found translates, in terms of the self-assembled membrane, into particles detaching away and eventually dissolving the membrane.

\subsection{Casimir-like interactions}\label{sec:casimir}

In the previous sections, we have adopted a mean-field approach. Here we wish to study whether equilibrium fluctuation-induced forces, i.e., Casimir-like interactions, can induce a transition between the unbound flat phase (U) and the bound flat phase (B). If this transition exists, it must be of first order since the two phases have the same symmetries. To isolate the Casimir effect, we take $C_0=0$ and $\kappa_b\to\infty$, so that the only effect of the adsorbed proteins is to locally stiffen the membrane. Because the protein has a zero spontaneous curvature, the SC phase will be absent. So, upon increasing $\mu$ we may either have a continuous increase of the protein density or a first-order phase transition.

Even in this simplified situation, it is very difficult to calculate exactly the free energy of the system for a given spatial distribution of proteins. We are  going to rely on estimates based the pointlike theory of Ref.~\cite{Dommersnes99EPJB}. In this work, the size $a$ of the protein inclusions is set by an upper wavevector cutoff, which allows to recover the results of Ref.~\cite{Goulian93EPL} for two extended inclusions. The multibody Casimir interaction is found to be exactly pairwise additive at leading order,  given by the sum of $-6T(a/R_{ij})^4$, where the $R_{ij}$'s are the distances between pairs of inclusions. Note that contrary to the results of Ref.~\cite{Weil10EPJE} for pinning inclusions, screening effects are very weak and occur only at the next orders.

At contact, i.e., for $R=2a$, which corresponds to the distance between nearest neighbors (NN), the above interaction gives $E_\mathrm{NN}\simeq-0.4\,T$. For $R=2\sqrt2a$, which corresponds to next nearest neighbors (NNN) in a square lattice, the interaction falls to $25\%$ of this value, while for $R=4a$, i.e., for second neighbors, it falls to $6\%$, which we will consider negligible. Note that these values should only be taken as estimates, since only the leading-order interaction has been taken into account, while at such short distances higher-order terms and multibody corrections are expected to play a significant role (as for curving inclusions~\cite{Fournier15EPJE}). A quick inspection of these corrections in the pointlike model revealed to us an increase in the attractiveness of the Casimir interaction.

\begin{figure}[h]
\centerline{\includegraphics[width=\columnwidth]{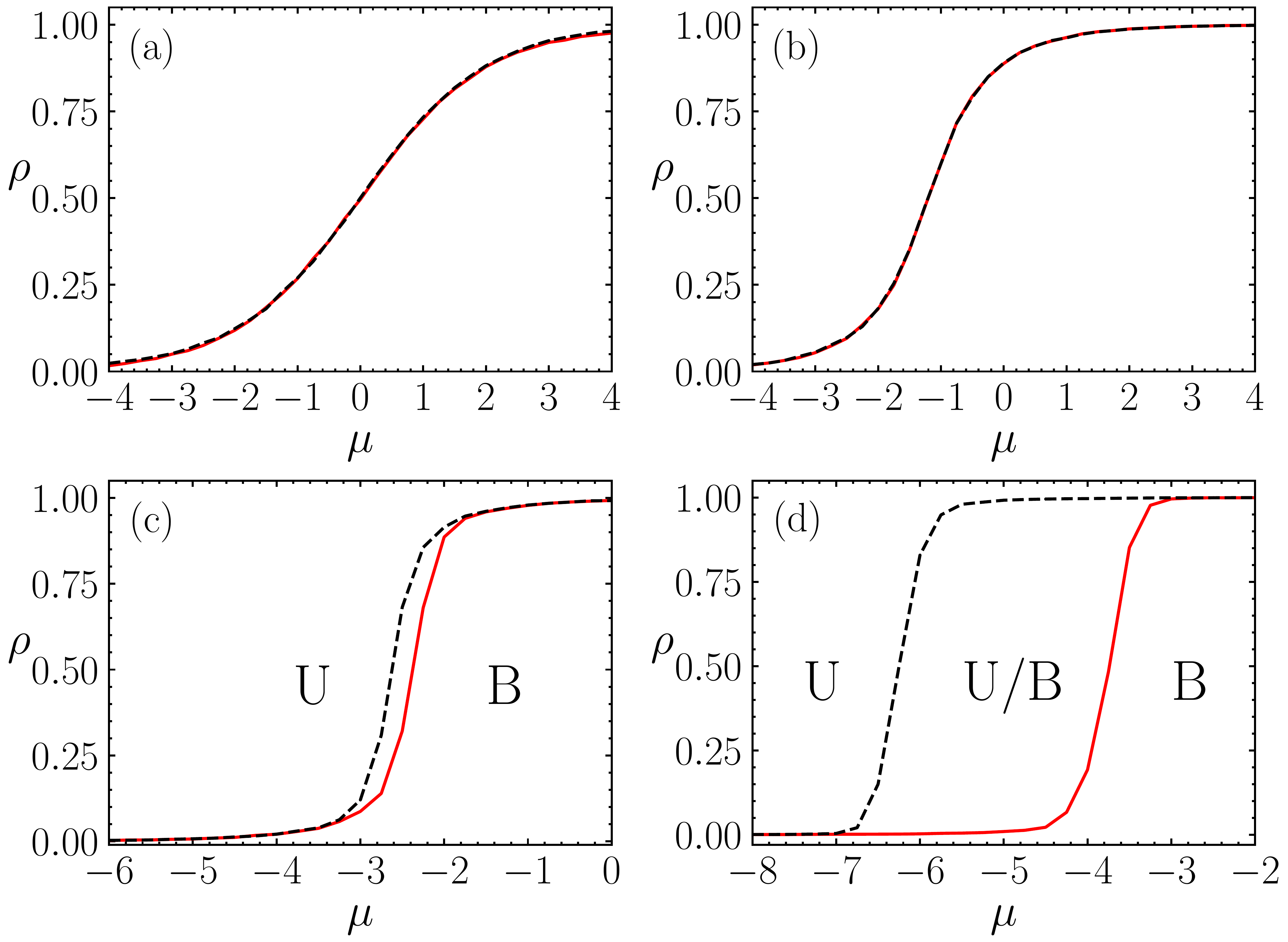}}
\caption{Equilibrium protein inclusion density $\rho$ versus chemical potential $\mu$ in a {MC} simulation of stiff membrane inclusions experiencing pure Casimir-like interactions (pairwise pointlike model). The solid red curve corresponds to an empty lattice initial condition. The dashed black curve corresponds to a fully occupied initial condition.
(a) $E_{\text{NN}}=E_{\text{NNN}}=0$.
(b) $E_{\text{NN}}=-0.5\,T$ and $E_{\text{NNN}}=-0.1\,T$.
(c) $E_{\text{NN}}=-T$ and $E_{\text{NNN}}=-0.25\,T$.
(d) $E_{\text{NN}}=-2\,T$ and $E_{\text{NNN}}=-0.5\,T$.
A first-order phase transition occurs in (c) and (d) as revealed by the coexistence between an unbound (U) and a bound  (B) state with different values of~$\rho$.}
\label{fig:Casi}
\end{figure}

To examine the effect of these Casimir interactions, we have performed a Monte Carlo {(MC)} simulation where particles diffuse on a square lattice, interact  through NN and NNN interactions only, and bind to the lattice, or detach from it by exchange with a reservoir of chemical potential $\mu$. The results, shown in Fig.~\ref{fig:Casi} indicate that the orders of magnitude given above are almost sufficient to produce an unbound-bound first-order phase transition.

\section{Simulation Model and Method}\label{sec:method}

A fluid membrane is numerically modeled in our work by a self-assembled one-layer sheet of $N$ particles.
The position and orientational vectors of the $i$-th particle are ${\bm{r}}_{i}$ and ${\bm{u}}_i$, respectively.
Since the details of this type of the meshless membrane model are described in Ref.~\citenum{shib11},
it is briefly described here.

The membrane particles interact with each other via a potential $U=U_{\mathrm {rep}}+U_{\mathrm {att}}+U_{\mathrm {bend}}+U_{\mathrm {tilt}}$.
The potential $U_{\mathrm {rep}}$ is an excluded volume interaction with diameter $\sigma$ for all pairs of particles.
The solvent is implicitly accounted for by an effective attractive potential  as follows:
\begin{eqnarray} \label{eq:U_att}
\frac{U_{\mathrm {att}}}{T} =  \frac{\varepsilon}{4}\sum_{i} \ln[1+\exp\{-4(\rho_i-\rho^*)\}]- C,
\end{eqnarray} 
where  $\rho_i= \sum_{j \ne i} f_{\mathrm {cut}}(r_{i,j})$, $C$ is a constant,
and $\rho^*$ is the characteristic density with $\rho^*=7$.
 In this study, we employ $\varepsilon=8$, which is greater than the values in our previous studies~\cite{shib11,nogu14},
to maintain the membrane in a wider parameter range.
$f_{\mathrm {cut}}(r)$ is a $C^{\infty}$ cutoff function~\cite{nogu06}
 and $r_{i,j}=|{\bf r}_{i,j}|$ with ${\bf r}_{i,j}={\bf r}_{i}-{\bf r}_j$:
\begin{equation} \label{eq:cutoff}
f_{\mathrm {cut}}(r)=\left\{ 
\begin{array}{ll}
\exp\{b(1+\frac{1}{(r/r_{\mathrm {cut}})^n -1})\}
& (r < r_{\mathrm {cut}}) \\
0  & (r \ge r_{\mathrm {cut}}) 
\end{array}
\right.
\end{equation}
where $n=6$, $b=\ln(2) \{(r_{\mathrm {cut}}/r_{\mathrm {att}})^n-1\}$,
$r_{\mathrm {att}}= 1.9\sigma$, and $r_{\mathrm {cut}}=2.4\sigma$. The set of parameters used above is described in detail in Ref.~\citenum{nogu14}.

The bending and tilt potentials
are given by 
 $U_{\mathrm {bend}}/T=(k_{\mathrm {bend}}/2) \sum_{i<j} ({\bm{u}}_{i} - {\bm{u}}_{j} - C_{\mathrm {bd}} \hat{\bm{r}}_{i,j} )^2 w_{\mathrm {cv}}(r_{i,j})$
and $U_{\mathrm {tilt}}/T=(k_{\mathrm{tilt}}/2) \sum_{i<j} [ ( {\bm{u}}_{i}\cdot \hat{\bm{r}}_{i,j})^2
 + ({\bm{u}}_{j}\cdot \hat{\bm{r}}_{i,j})^2  ] w_{\mathrm {cv}}(r_{i,j})$,  respectively,
where 
 $\hat{\bm{r}}_{i,j}={\bm{r}}_{i,j}/r_{i,j}$ and $w_{\mathrm {cv}}(r_{i,j})$ is a weight function. The energy $U_{\mathrm {bend}}$ penalizes the splay of ${\bm u}_i$, and hence penalizes the curvature of the membrane when $C_{\mathrm {bd}} = 0$, while it favors a spontaneous curvature $C_0 = C_{\mathrm {bd}}/(2 \sigma)$ when $C_{\mathrm {bd}}$ is nonzero~\cite{shib11}. As for $U_{\mathrm {tilt}}$, it penalizes the lipid tilt, therefore favors the normal orientation of the lipids relative to the membrane plane.

The membrane consisting of $25~600$ membrane particles is simulated under periodic boundary conditions
with $N\gamma T$ ensemble, where $\gamma$ is the surface tension
that is conjugate to the projected area onto the $xy$ plane as defined in Section~\ref{sec:theory}.
The projected area of the square membrane ($A_p=L_p^2$) is a fluctuating quantity~\cite{fell95,nogu12}.
The motion of the particle position ${\bf r}_{i}$ and 
the orientation ${\bf u}_{i}$ are given by underdamped Langevin equations,
which are integrated by the leapfrog algorithm \cite{alle87,nogu11}
with $\Delta t=0.002\tau_0$ where $\tau_0= \sigma^2/D_0$ for the time unit,
where $D_0$ is the diffusion coefficient of the free membrane particles.

Here, each membrane particle is a binding site and can be found in two---bound or unbound---states.
In this study, $C_0=0$ and $k_{\mathrm {bend}}=k_{\mathrm{tilt}}=10$ for the unbound membrane particles
and $k_{\mathrm {bend}}=k_{\mathrm{tilt}}=80$ for the bound membrane particles, in which
 $\kappa_u/T=16 \pm 1$ and $\kappa_b/T=144 \pm 7$.
In the bending and tilt potentials, for a pair of neighboring bound and unbound particles,
we use the mean value {$k_{\mathrm {bend}}=k_{\mathrm{tilt}}=45$}.
We find that the ratio of the Gaussian modulus $\bar{\kappa}$ to $\kappa$ is uniform, independently of the local binding fraction~\cite{nogu19}: 
$\bar{\kappa}/\kappa=-0.9\pm 0.1$.
\textcolor{black}{
In the following, we shall mainly consider tensionless membranes and membranes with tension $\gamma=0.5T/\sigma^2$. With typically $\sigma\approx 10$\,nm and $T\approx 4\times10^{-21}$\,J, this corresponds to an average tension $\approx  0.02$\,mN/m, well below usual lysis tensions ($1$--$25$\,mN/m)~\cite{evan00,evan03,ly04}.
}
For the unbound particles, the membrane area per particle is
 $1.251\sigma^2$ and $1.257\sigma^2$ at $\gamma=0$ and $0.5T/\sigma^2$, respectively. It is slightly larger (by a few percents) for bound particles:
$1.294\sigma^2$ and $1.300\sigma^2$ at $\gamma=0$ and $0.5T/\sigma^2$, for $C_0\sigma=0.1$.
The unit length $a$ of the theory (governing the surface area $a^2$ covered by a bound protein, defined in Section~\ref{sec:theory}) is thus found to be $a \simeq 1.1\sigma$.

The bound and unbound states are stochastically switched by a Metropolis {MC} procedure with the acceptance rate $p_{\mathrm {acpt}}$:
\begin{equation}\label{eq:Metro}
p_{\mathrm {acpt}} = \left\{
\begin{array}{ll}
\exp(\pm \Delta H/T ) &{\mathrm {if\ }} \pm \Delta H <0, \\
  1 &{\mathrm {otherwise}},
\end{array}
\right.
\end{equation}
where the $+$ and $-$ signs refer to the unbinding and binding processes, respectively.
Here, $\Delta H= \Delta U - \mu$ where $\Delta U$ is the energy difference between the bound and unbound states
and $\mu$ is the chemical potential of particles attempting to bind.
We also consider active unbinding in which the particles change from bound to unbound states with a rate $\eta_2$ independent of the state of the system, as defined in eqn~(\ref{eq:cartoon-activity}). Our simulations are carried out at $\eta_1=0$, \textcolor{black}{otherwise membrane particles spontaneously detach, because active binding can lead to a higher bending energy than the attractive energy between membrane particles}.
In thermal equilibrium ($\eta_2=0$), static properties are independent of the rates of the binding/unbinding processes. Out of equilibrium, however, the steady state reached by the system {\it a priori} depends on the details of the binding and unbinding processes (compared with the thermal binding-unbinding rates and with the membrane dynamics), as long as detailed-balance is broken. Here, we choose to consider relatively fast binding/unbinding processes compared to the membrane motion.
For each membrane particle, the Metropolis MC and active unbinding processes are performed every $\tau_{\mathrm b}=0.01\tau_0$
with probabilities  $\alpha_{\mathrm {mp}}\tau_{\mathrm b}$  and $\eta_2\tau_{\mathrm b}$ respectively.
We fix $\alpha_{\mathrm {mp}}\tau_0= 10$ and vary the ratio $\eta_2/\alpha_{\mathrm {mp}}$ in our investigation.
The binding fraction is essentially controlled by the ratio  $\eta_2/\alpha_{\mathrm {mp}}$,
as the binding/unbinding is faster than membrane deformation.  We have checked it by varying $\alpha_{\mathrm {mp}}\tau_0$ at $C_0\sigma= 0.1$, $\gamma=0.5T/\sigma^2$, and $\mu=6T$.

In our analysis, in order to characterize the various phases we find, we resort to the concept of cluster. Two sites are considered to belong to the same cluster
when the distance between them
is less than $r_{\mathrm {att}}$. 
The probability $P(i_{\mathrm {cl}})$ that a site belongs to a cluster of size $i_{\mathrm {cl}}$
is  $P(i_{\mathrm {cl}})= \langle n_{\mathrm {cl}} i_{\mathrm {cl}} \rangle/N$, 
where $n_{\mathrm {cl}}$ is the number of clusters consisting of $i_{\mathrm {cl}}$ sites.
The vertical span of the membrane is calculated from 
the membrane height variance as 
$z_{\mathrm {mb}}^2=\sum_{i}^{N} (z_i-z_{\mathrm G})^2/N$,
where $z_{\mathrm G}=\sum_{i}^{N} z_i/N$.

The results of the simulations are normalized by the particle diameter $\sigma$, by  temperature $T$ and by the time step $\tau_0$ for lengths, energies and times, respectively.
The subscripts $b$ and $u$ indicate bound and unbound states, while {U, B, and SC} refer to the unbound, bound and separated-corrugated phases, respectively.
The error bars show the standard deviation calculated from three or more independent runs.

\section{Simulation Results}\label{sec:results}

\subsection{Phase Separation in Thermal Equilibrium}\label{sec:thermal}

\begin{figure}[]\centering
\includegraphics[width=7cm]{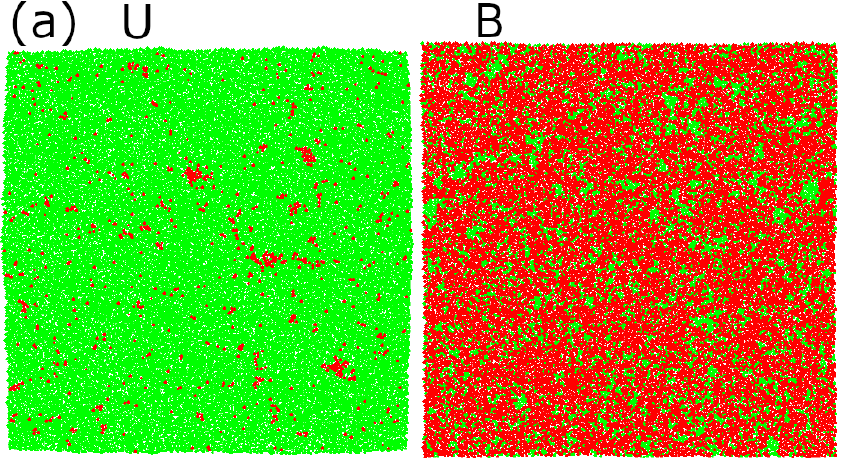}
\includegraphics[width=7cm]{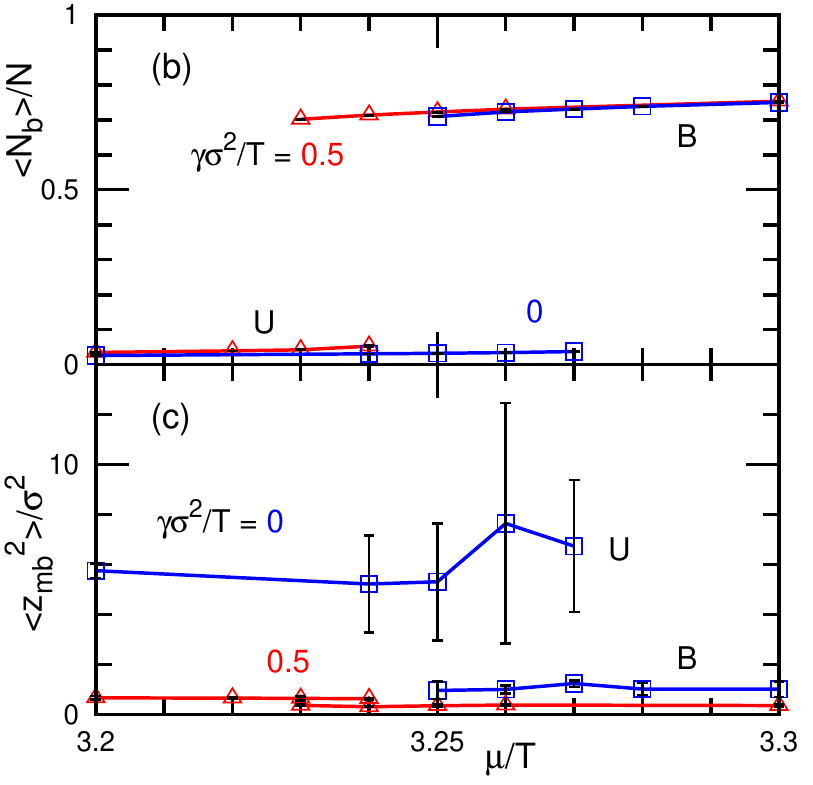}
\caption{
Discontinuous binding transition at $C_0= 0$ in thermal equilibrium ($\eta_2=0$).
(a)  Snapshots of the unbound (U) and bound (B) phases existing at the same chemical potential $\mu/T=3.26$ for $\gamma=0$.
The bound and unbound sites are displayed in red 
and in  green, respectively.
(b) Binding density $\langle N_b\rangle/N$
and (c) vertical membrane span $z_{\mathrm {mb}}$
as a function of the chemical potential $\mu$  at  $\gamma\sigma^2/T=0$ and $0.5$.
}
\label{fig:c00}
\end{figure}

We now describe the membrane behavior in thermal equilibrium (without the active unbinding, at $\eta_2=0$). First, we work at $C_0=0$, so that the only action of the bound and unbound particles is to locally alter the bending rigidity. As the chemical potential $\mu$ increases,
the membrane exhibits a discontinuous transition from the unbound state (U) to the bound state (B) (see Fig.~\ref{fig:c00}).
The two states can exist at the same chemical potential in the vicinity of the phase transition. 
This transition occurs due to the suppression of membrane fluctuations by the high bending rigidity of the bound sites. It is therefore Casimir interactions that drive this transition, as discussed in Section~\ref{sec:casimir}.
{As the surface tension $\gamma$ increases, the transition occurs at slightly lower values of $\mu$
and the coexistence range becomes narrower.}

At this stage, we would like to compare with the theoretical results shown in Fig.~\ref{fig:Casi}. In both approaches, a first-order transition is observed. Because surface tension flattens out the membrane by stretching it, we expect that Casimir interactions will be all the weaker as surface tension {increases}.

\begin{figure}\centering
\includegraphics[width=.9\columnwidth]{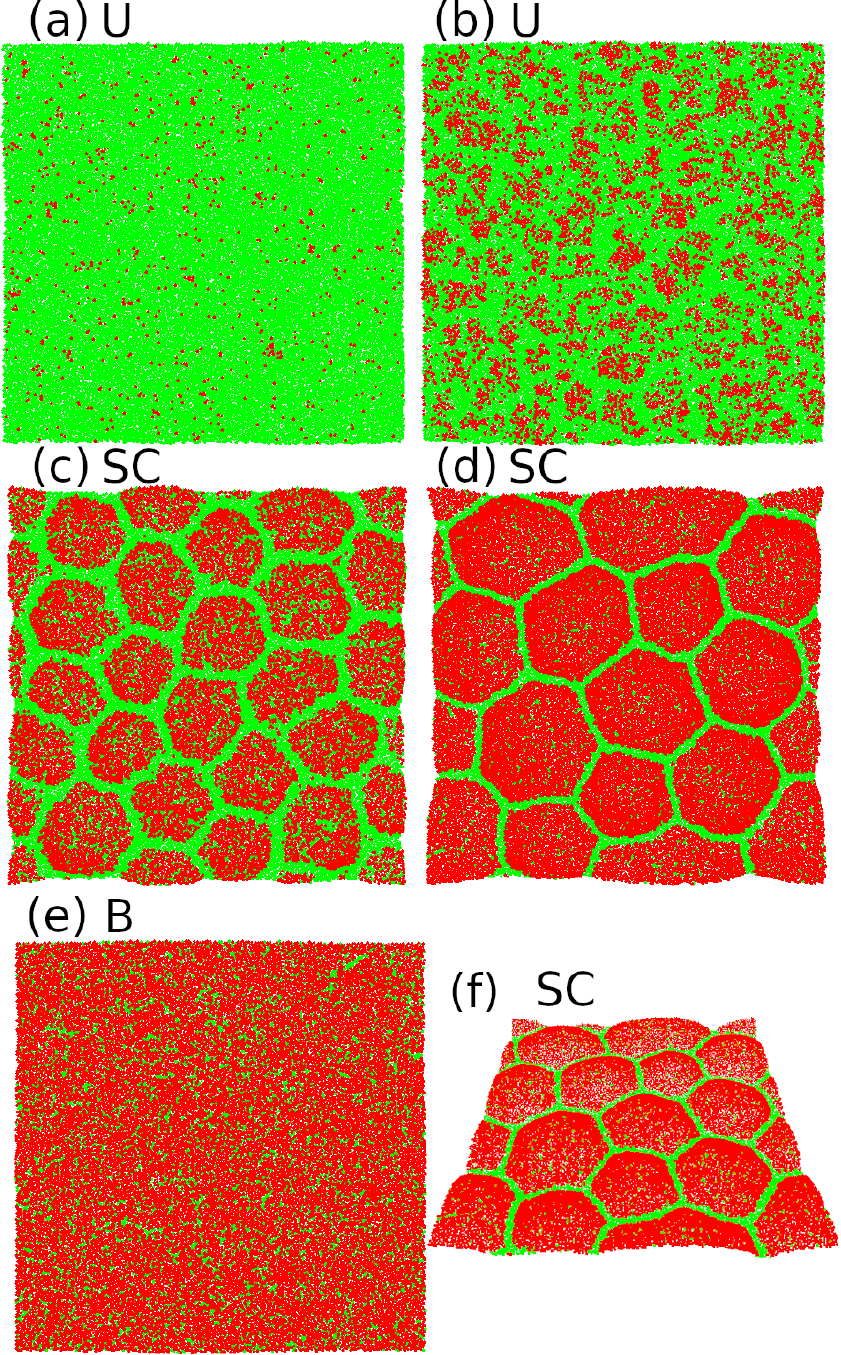}
\caption{
 Snapshots of membranes at $C_0\sigma= 0.1$, $\gamma\sigma^2/T=0.5$, and $\eta_2=0$.
(a) U phase at $\mu/T=4$, (b) Close to the phase boundary between U and SC at $\mu/T=5$, 
(c) SC phase at $\mu/T=6$. (d)--(f)  SC and B phases at $\mu/T=8$.
The top views are shown in (a)--(e) and
a bird's eye view of the snapshot of (d) is shown in (f). 
}
\label{fig:snapeq}
\end{figure}

\begin{figure}[]\centering
\includegraphics[width=.9\columnwidth]{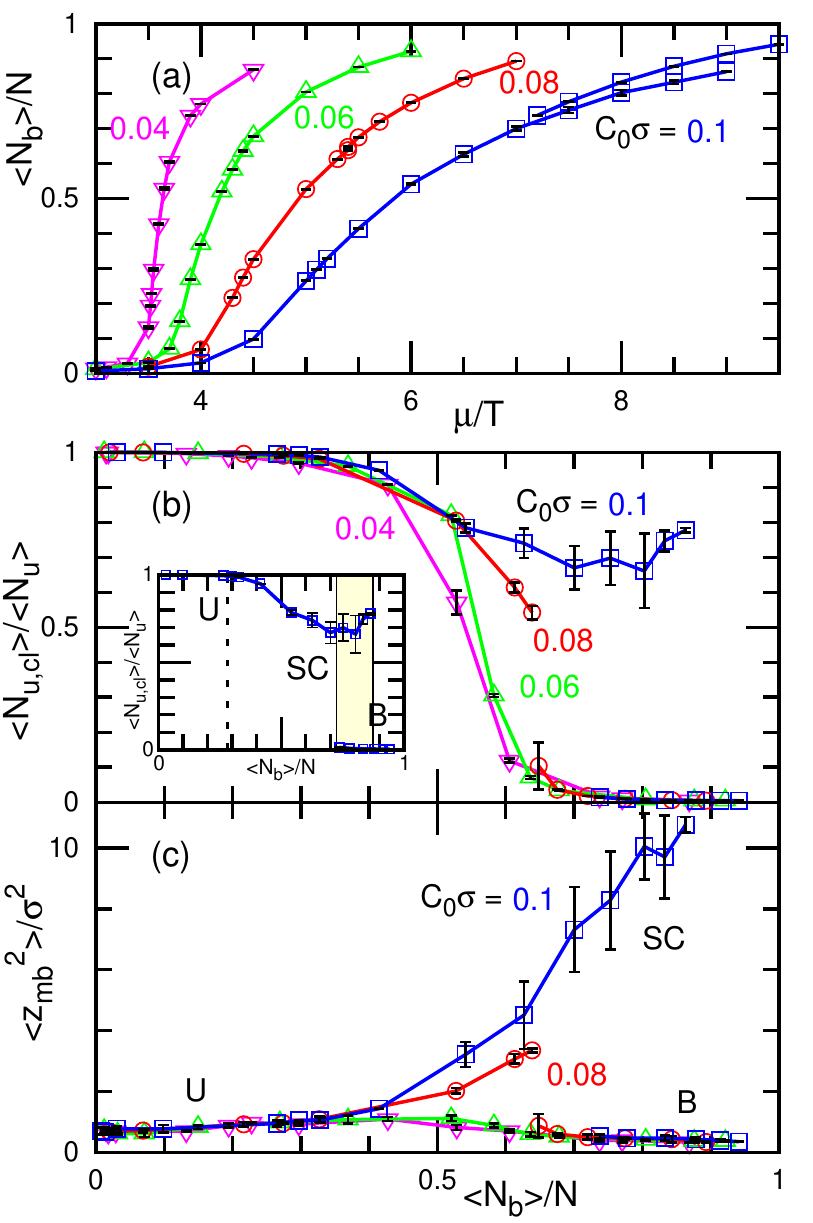}
\caption{
Binding  at $\gamma\sigma^2/T=0.5$ and $\eta_2=0$.
(a) binding density $\langle N_b\rangle/N$.
(b) ratio of the largest cluster of unbound sites $\langle N_{u,{\mathrm {cl}}}\rangle/\langle N_u\rangle$. As shown for $C_0\sigma=0.1$, the U, SC and B phases can be determined by the value of $\langle N_{u,{\mathrm {cl}}}\rangle/\langle N_u\rangle$. The dashed line corresponds to the transition between the U and SC phases and the yellow domain corresponds to the region where the SC and B phases coexist. (c) vertical membrane span $z_{\mathrm {mb}}$. 
}
\label{fig:eq0}
\end{figure}

\begin{figure}[]\centering
\includegraphics[width=.9\columnwidth]{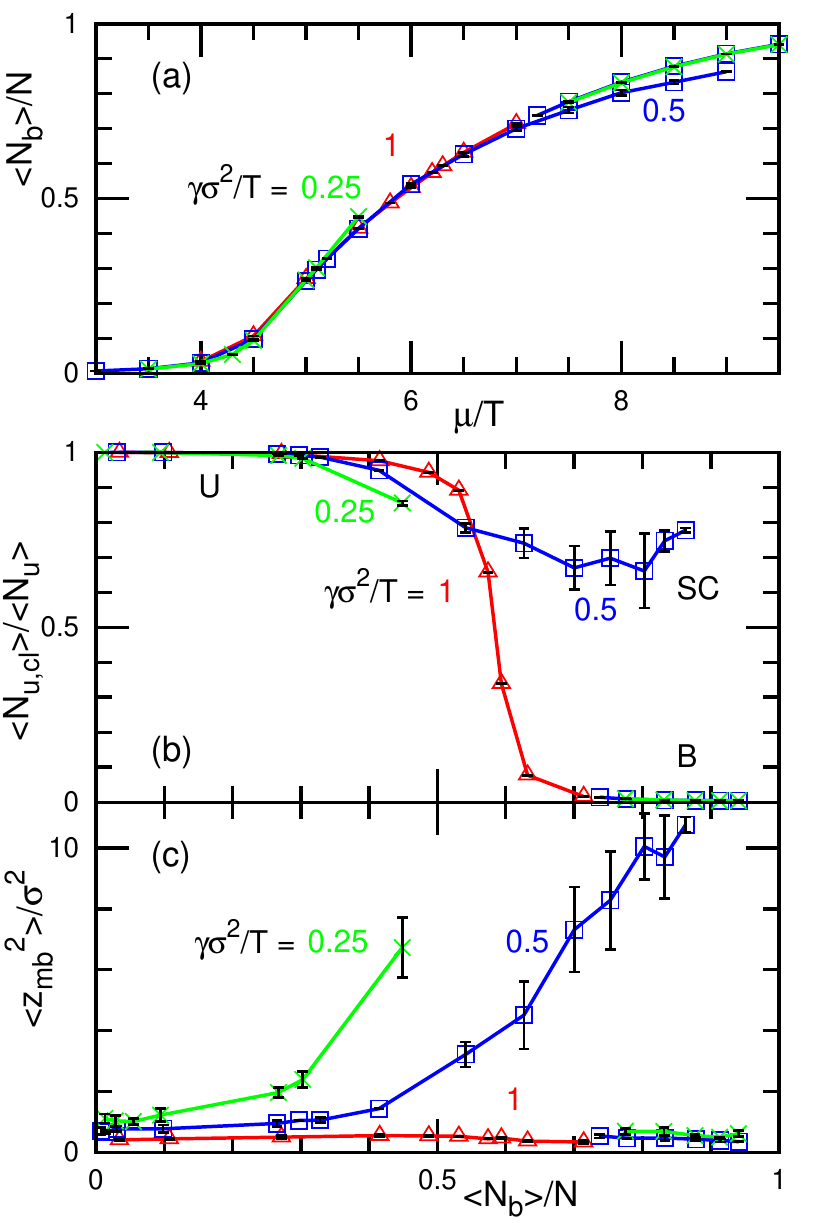}
\caption{
Surface tension $\gamma$ dependence at $C_0\sigma= 0.1$ and $\eta_2=0$.
(a) Binding density $\langle N_b\rangle/N$.
(b) ratio of the largest cluster of unbound sites $\langle N_{u,\mathrm{cl}}\rangle/\langle N_u\rangle$. (c) vertical membrane span $z_{\mathrm {mb}}$.
At $\gamma\sigma^2/T=0.25$, the SC phase is unstable for $\mu/T\gtrsim 6$.
}
\label{fig:ten}
\end{figure}

\begin{figure}\centering
\includegraphics{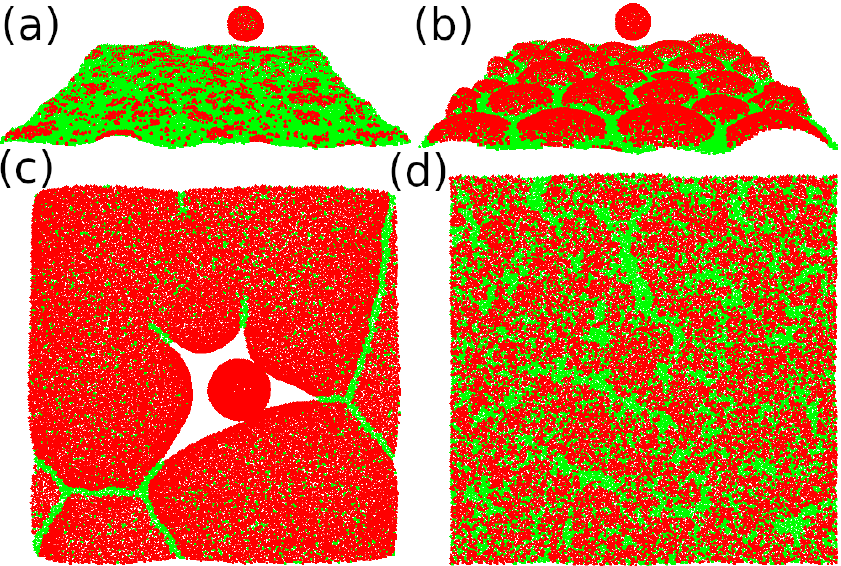}
\caption{
 Snapshots of membranes at $\eta_2=0$.
 (a),(b) Vesicle formation (a) at  $C_0\sigma= 0.1$, $\gamma=0$, and $\mu/T=4.7$
 and (b) at $C_0\sigma= 0.1$, $\gamma\sigma^2/T=0.25$, and  $\mu/T=6$.
 (c) Membrane rupture at  $C_0\sigma= 0.1$, $\gamma\sigma^2/T=0.25$, and $\mu/T=8.5$.
 (d) Anisotropic clusters of unbound particles at $C_0\sigma= 0.06$, $\gamma\sigma^2/T=0.5$, and $\mu/T=4.3$.
Bird's eye and top views are shown in (a),(b) and in (c),(d), respectively.
}
\label{fig:snapves}
\end{figure}

\begin{figure}\centering
\includegraphics{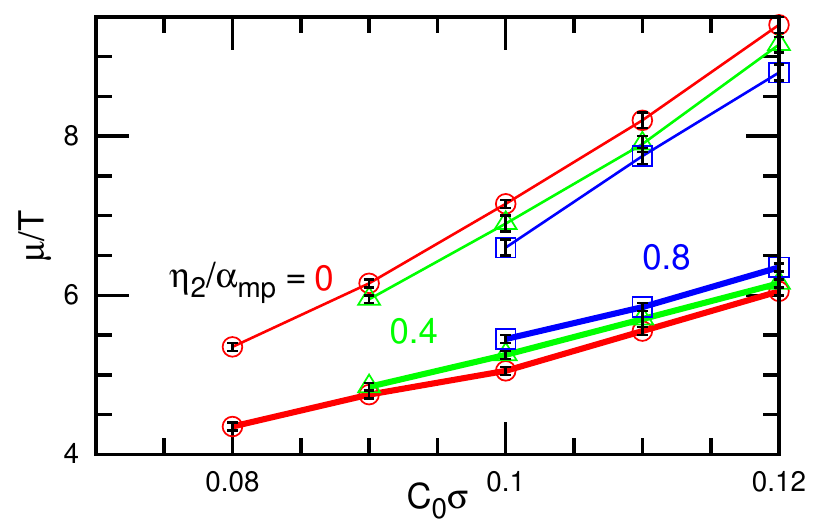}
\caption{
  Phase boundaries for the metastability of the unbound (lower branch) and bound (upper branch) states
  for $\eta_2/\alpha_{\mathrm {mp}}=0$ ($\circ$), $0.4$ ($\triangle$), and $0.8$ ($\Box$) at $\gamma\sigma^2/T=0.5$.
  The unbound and bound states exist as stable or metastable states
  in the regions below the lower line and above the upper line, respectively.
  The membrane is phase-separated in the region between two lines. 
{Note that the membrane continuously changes from the U to B phases at $\mu/T=0.06$.} We conjecture that the red branches eventually merge at around $C_0\sigma\simeq 0.06$ and continue into a line for lower values of $C_0\sigma$.
}
\label{fig:pd}
\end{figure}

\begin{figure}\centering
\includegraphics{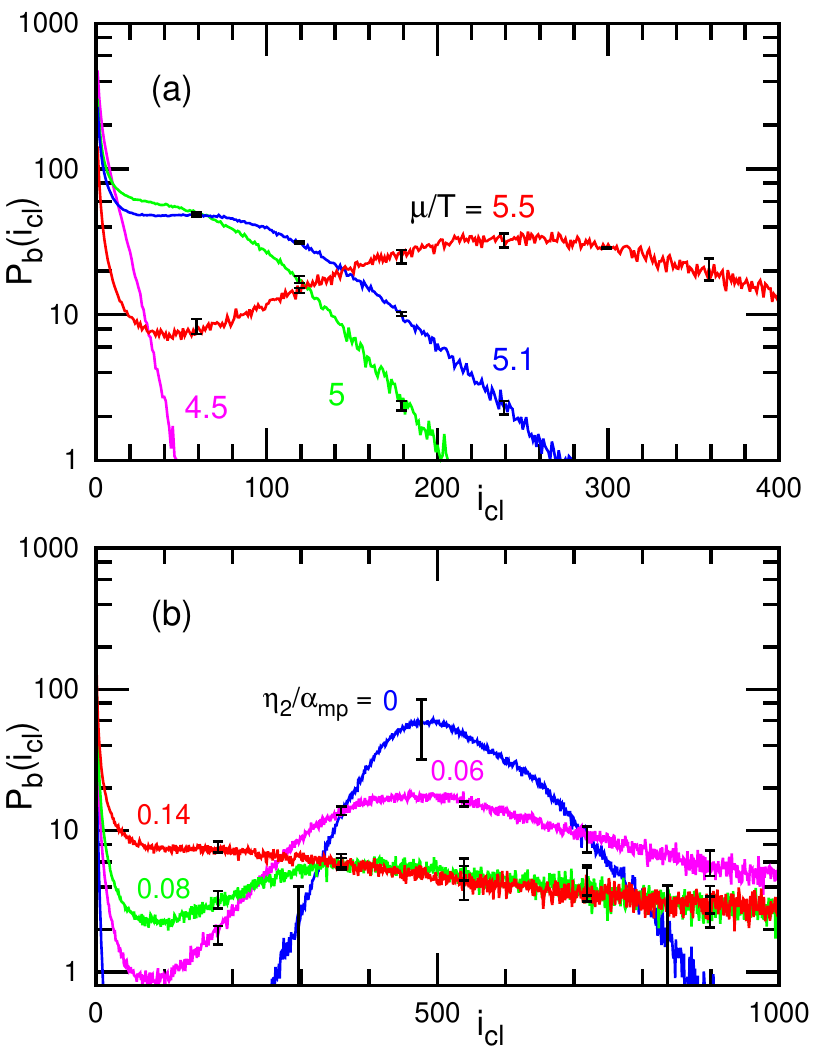}
\caption{
Size distribution $P_b(i_{\mathrm {cl}})$ of bound-site cluster at $C_0\sigma= 0.1$ and $\gamma\sigma^2/T=0.5$.
(a) $\mu/T=4.5$, $5$, $5.1$, and $5.5$ at $\eta_2=0$.
(b) $\eta_2/\alpha_{\mathrm {mp}}=0$, $0.6$, $0.8$, and $1.4$ at $\mu/T=6$.
}
\label{fig:cnhis}
\end{figure}

For the finite spontaneous curvatures ($C_0 \ne 0$),
the bound sites prefer to assemble to a curved domain leading to the formation of spatial patterns
(Figs.~\ref{fig:snapeq}--\ref{fig:cnhis}).
At a high spontaneous curvature ($C_0\sigma = 0.1$) and a medium surface tension ($\gamma\sigma^2/T=0.5$),
an SC state, where micro-domains of bound sites are formed, appears
between the U and B phases (see Fig.~\ref{fig:snapeq}).
To maintain a flat membrane in average under the periodic boundary condition,
the bound sites form finite-size bowl-shaped domains, and the unbound membrane between the bound domains
is bent in the opposite direction [see Figs.~\ref{fig:snapeq}(f)].
When the membrane is sliced along a vertical plane, the cross section has a bump shape as depicted in Fig.~\ref{fig:hump}.
A hexagonal-shaped pattern is formed instead of the 1D bump pattern, since a spherical shape is preferred by the isotropic spontaneous curvature.
However, the essential feature of the stability is captured in the 1D shape.
\textcolor{black}{Similar curved domains can be formed on protein-free membranes. Such curve-shaped domains were observed in three-component lipid vesicles~\cite{baum03,veat03,yana08,chri09}. These domains are, however, caused by their strong line tension unlike our case (we have almost no line tension as there is no direct repulsion between bound and unbound particles). }

Examples of the initial relaxation dynamics at $\mu/T=7$ are shown in Movies 1 and 2 \textcolor{black}{provided in the ESI}.
The unbound domains elongately grow leading to a percolated network.
When initial states are set to the smaller or larger domains obtained at low or high $\mu$,
the domains grow or are reduced, respectively, but do not completely converge to the same size
even in the long simulation runs by a hysteresis. 
In the SC states, the simulations are performed from several different initial states to check this hysteresis.
The error bars in Figs.~\ref{fig:eq0}, \ref{fig:ten}, and \ref{fig:ac1} show the width of the obtained values due to the hysteresis.
The bound domain size increases with increasing $\mu$ [compare Figs.~\ref{fig:snapeq}(c) and (d)].

At $C_0\sigma = 0.1$ and $\gamma\sigma^2/T=0.5$, the transition between the U and SC phases is continuous,
but the transition between the SC and B phases is discontinuous
and two phases coexist around the phase boundary [see Figs.~\ref{fig:snapeq}(d)--(f) and \ref{fig:eq0}(a)].
This agrees with the theoretical prediction in Section~\ref{sec:naeq}.
To detect the discreteness of the transition between the  SC and B states more clearly,
the ratio of the average number of sites in the largest unbound cluster to that in the unbound sites,
$\langle N_{u,\mathrm{cl}} \rangle/\langle N_u \rangle$, is shown 
with respect to $\langle N_b \rangle/N$ in Fig.~\ref{fig:eq0}(b).
This ratio is close to unity in the SC phase and close to null in the B phase state, because of the unbound region percolates.
A large gap exists between the two states at $C_0\sigma=0.08$ and $0.1$.
In the SC state, most of the unbound sites belong to the largest cluster
so that the unbound sites form a single percolated domain.
The mean membrane vertical span $\langle z_{\mathrm {mb}}^2\rangle$ also exhibits a discrete gap [see Fig.~
\ref{fig:eq0}(c)],
since the SC membranes are largely bent whereas not in the B phase.

For the low spontaneous curvatures ($C_0\sigma = 0.04$ and $0.06$ at $\gamma\sigma^2/T=0.5$) or high surface tension ($\gamma\sigma^2/T=1$ at $C_0\sigma = 0.1$),
the transition between the SC and B phases becomes continuous [see Fig.~\ref{fig:eq0} and \ref{fig:ten}].
In the SC phase, the unbound domains are of anisotropic shapes but {do not} form a fixed network structure
so that stable micro-domains are not formed [see Fig.~\ref{fig:snapves}(d) and corresponding Movies 3 \textcolor{black}{provided in the ESI}].
This corresponds to the single phase at $C_0<C_{\mathrm {th}}$ in the theory,
where the binding ratio gradually changes.
For zero or low surface tension ($\gamma\sigma^2/T=0$ or $0.25$),
the micro-domains of the bound sites form vesicles via budding [Figs.~\ref{fig:snapves}(a),(b)] or membrane rupture [Fig.~\ref{fig:snapves}(c)].
Therefore, finite tension is required to stabilize the SC phase.

As theoretically analyzed in Fig.~\ref{fig:eq_ph_diag},
the phase boundary of the uniform (unbound and bound) phases is shown in Fig.~\ref{fig:pd}.
In the region between two lines, the uniform phase does not exist even as a metastable state.
At $\eta_2 =0$, the unstable region becomes wider at higher values of $C_0$.
The lower boundary is determined by the appearance of a peak in the cluster size distribution $P_b(i_{\mathrm {cl}})$ of the bound sites,
as shown in Fig.~\ref{fig:cnhis}(a).

\begin{figure}\centering
\includegraphics{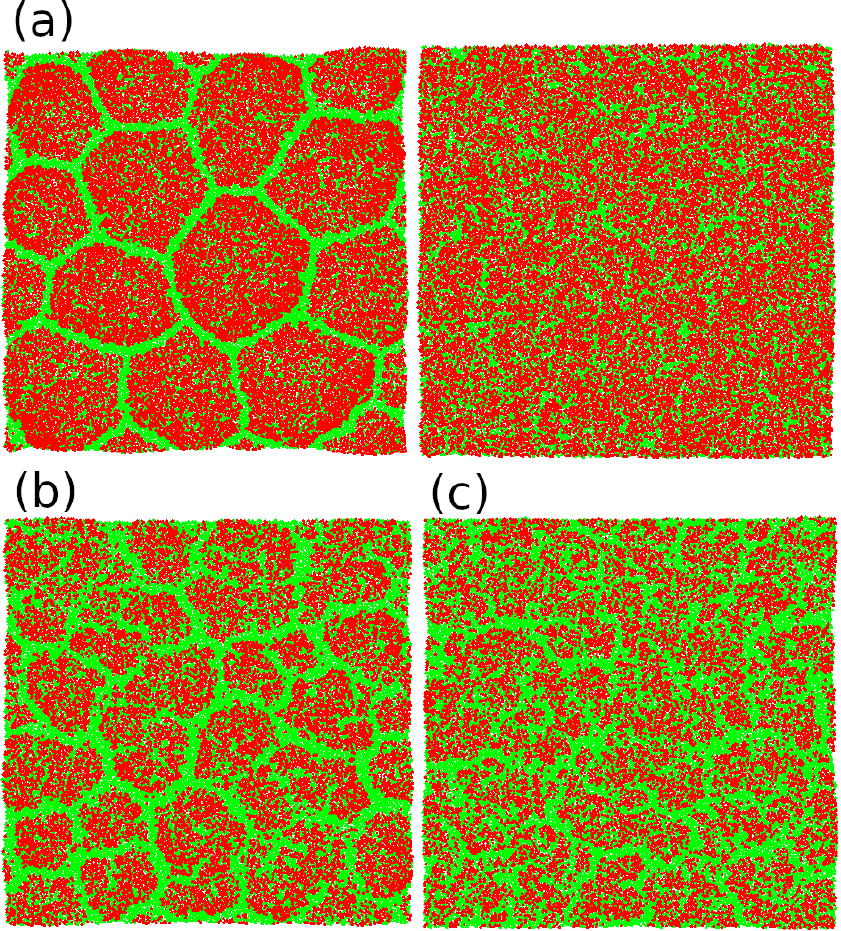}
\caption{
 Snapshots of membranes with active unbinding at $C_0\sigma=0.1$ and $\gamma\sigma^2/T=0.5$
for (a) $\{\mu/T,\eta_2/\alpha_{\mathrm {mp}}\}=\{7,0.08\}$,
(b) $\{6,0.06\}$, and (c) $\{6,0.12\}$.
}
\label{fig:ac0}
\end{figure}

\subsection{Phase Separation out of equilibrium}\label{sec:active}

\begin{figure}\centering
\includegraphics[width=.9\columnwidth]{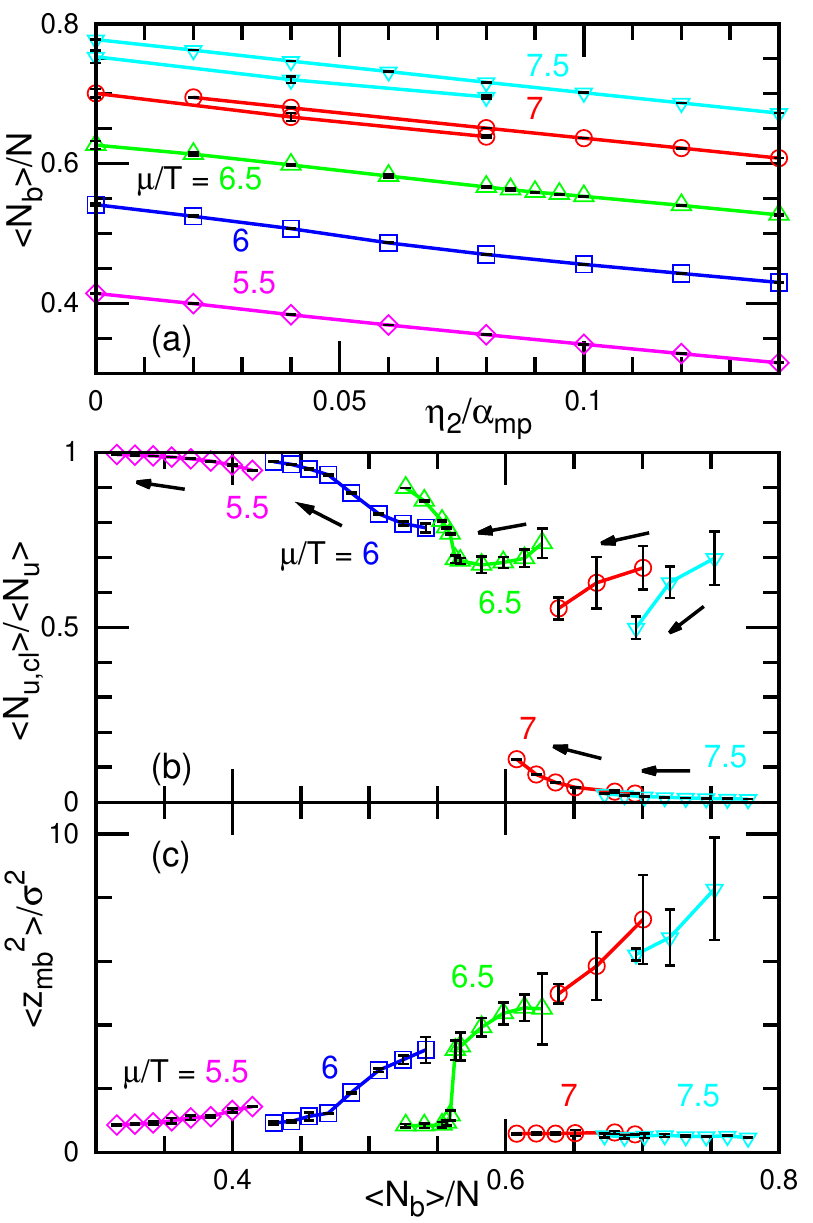}
\caption{
Membrane with active unbinding at $C_0\sigma=0.1$ and $\gamma\sigma^2/T=0.5$.
(a) Binding density $\langle N_b\rangle/N$ as a function of the active unbinding rate $\eta_2$.
(b) Ratio of the largest cluster of (a) unbound sites $\langle N_{u,\mathrm{cl}}\rangle/\langle N_u\rangle$
as a function of the binding density $\langle N_b\rangle/N$. The arrows show the direction of increasing $\eta_2$.
(c) Vertical membrane span $z_{\mathrm {mb}}$. 
}
\label{fig:ac1}
\end{figure}

In this section, we describe the membrane behavior in the nonequilibrium regime with active unbinding ($\eta_2 > 0$).
As $\eta_2$ increases, the ratio of the bound sites linearly decreases, 
and the bound and unbound sites are mixed more randomly as shown in Figs.~\ref{fig:ac0} and \ref{fig:ac1}(a).
From Fig.~\ref{fig:ac1}(b), we see that the SC phase becomes unstable and eventually disappears as $\eta_2$ is increased at not too large chemical potential. This property is also apparent in Fig.~\ref{fig:pd} where the lower boundary rises upwards. 
As the fraction of bound sites decreases with increasing $\eta_2$,
the domain structure in the SC phase becomes less pronounced (Fig.~\ref{fig:ac0}),
and the membrane adopts a flatter shape (Fig.~\ref{fig:ac1}(c)).
The SC and uniform phases are clearly distinguished from the fraction of unbound sites belonging to the largest cluster 
at $\mu/T=7$ and $7.5$, while, as $\eta_2$ is increased, the SC phase is continuously blurred out when $\mu/T \lesssim 6.5$ (Figs.~\ref{fig:ac1}(b)).
With increasing $\eta_2$,  $P_b(i_{\mathrm {cl}})$ has a lower and broader peak, and 
subsequently, it monotonously decreases (see Fig.~\ref{fig:cnhis}(b)).
This is different from the transition between the U and SC phases in equilibrium, where large clusters are exponentially rare in the U phase, as shown in  Fig.~\ref{fig:cnhis}(a). 

In the theoretical analysis of Section~\ref{sec:theory} we observe (Fig.~\ref{fig:activeLSA}(a)) that a nonzero $\eta_2$ shifts the lower boundary of yellow metastability region upwards, in agreement with the simulation results in Fig.~\ref{fig:pd}, but it also shifts its upper boundary upwards, which is not consistent with the observed numerics (Fig.~\ref{fig:pd}). Such nonequilibrium features as the increased fuzziness cannot be accounted for within the framework of a linear stability analysis.

\subsection{Discussion}\label{sec:dis}

The aforementioned theoretical and simulation results  agree qualitatively well,
but some differences are seen. We now discuss these in more detail.
In comparing the phase diagrams in Figs.~\ref{fig:eq_ph_diag} and \ref{fig:pd},
the chemical potential $\mu$ for the SC phase is roughly $5T$ higher in the simulation than $\mu$ in the theory,
but the range of the SC phase is compatible.
In the theory, the binding changes only the bending energy.
Conversely, in the simulation, it also modifies the other energy ($U_{\mathrm {rep}}, U_{\mathrm {att}}$)
and the local membrane area is slightly changed by the binding.
Hence, this shift of $\mu$ might be caused by this different energy change so that
 higher $\mu$ is required for binding to occur in the simulation.
On the other hand, the differences of the typical threshold values of $C_0$ in the phase diagrams are small.
Part of these differences are due to the differences in the length units ($a\simeq 1.1\sigma$),
and the rest are likely due to thermal fluctuations. Indeed, the phase boundaries can be affected by thermal fluctuations.

In the analytical approach, a first-order transition between U and SC phases is predicted in the vicinity of the critical point ($C_0<C_{\mathrm {th}}$),
as shown in Fig.~\ref{fig:eq_ph_diag}.
By contrast, this transition is always found to be continuous in the simulation.
Since this appears only in the vicinity of  the critical point,
the free-energy barrier between the two states is presumably small.
Thermal fluctuations may smear out the free-energy barrier in the simulation.

The active unbinding shrinks the region of the SC phase in the phase diagram of the simulation (see Fig.~\ref{fig:pd}).
By contrast, a shift to higher $\mu$ is predicted in the analytical model (see Fig.~\ref{fig:activeLSA}(a)).
This difference may be due to the setting of the binding/unbinding using different choices for the rates $\alpha_{1,2}$ and $\alpha_{\rm mp}p_\text{acpt}$ in the continuum equation  and in the MC method.

\section{Outlook}\label{sec:sum}

In this work, we have studied the structuring of membranes interacting with binding molecules that locally constrain the membrane curvature and increase its bending rigidity. Our analysis has relied on simulations of a meshless membrane model and on an analytical coarse-grained description of the dynamics (based on the Helfrich energy). 

In thermal equilibrium, without any active binding/unbinding, we have found that for high spontaneous curvatures and intermediate densities, bound sites locally self-assemble into a bowl-like shape. The membrane then exhibits, at the macroscopic scale,  a hexagonally corrugated shape (SC phase). At low density or high density, the membrane adopts a flat state with a uniform distribution of particles (unbound or bound phase, respectively). Second and first-order transitions occur between the SC and unbound phases and between the SC and bound phases, respectively. For a small nonzero spontaneous curvature or under high surface tension, the density of bound sites gradually increases from the completely unbound state up to the bound state as the chemical potential is increased. At zero spontaneous curvature, we have found that Casimir-like interactions induce a first-order transition from the unbound to bound states. 
Both analytical approaches and simulations agree with each other. 

Out of equilibrium, in the presence of an active binding or unbinding, our analytical analysis, based on Glauber equilibrium transition rates, predicts a shift of the SC phase towards higher spontaneous curvatures, as observed in the simulations.
It also predicts a shift of the SC phase to lower chemical potentials in the active binding case and to higher chemical potentials in the active unbinding case, while the simulations show a simple shrinkage of the SC phase in the latter case.
We expect this discrepancy to be due to the difference in the implementation of the equilibrium transition rates (Glauber vs Metropolis).
Simulations show that active unbinding makes the bowl-shaped domains of the SC phase fuzzier. We observe that the  SC phase disappears at small curvatures, and because we have a discontinuous transition between the U and B phases at zero curvature, we conjecture the closing of the SC phase is continued by a line connecting to that zero curvature point.

Casimir interactions are the result of thermal fluctuations. Because of their attractive nature, we expect a difference between simulations (that take these forces into account) and the nonlinear analysis which neglects fluctuations. Casimir forces act as a stabilizing force for the SC phase, which should thus appear for a broader range of parameters in the simulation than in the analysis of the mean-field PDEs.

We have assumed the thermal and active binding/unbinding rates to be time-independent.
This means that the protein diffusion in the bulk is faster than the binding/unbinding
and the protein concentration of the bulk in the vicinity of the membrane surface is maintained constant. However, it is known that 
large proteins exhibit slow diffusion and the lowering of the density of proteins in the vicinity of the membrane suppresses the binding.
When the binding/unbinding is compatible and faster, the dynamics of the protein in the bulk might also play an important role in determining the nonequilibrium steady state,
in which a convection flow may be generated. And in fact, a more realistic description should also incorporate the combined hydrodynamics of the membrane and of the solvent and the diffusion/convection of the bulk particles.

Finally, traveling waves were reported in other systems both in 1D
geometry~\cite{cagn18} and in 2D geometry~\cite{zaki18}. These traveling waves are a rather general feature of the reaction--diffusion dynamics coupled with membrane deformation~\cite{pele11,wu18,tame20}. In this study, we have modeled the activity by a constant rate process. More complex dependence of the latter rate, {\it e.g.} including feedback from the other dynamical fields, may thus lead to such complex structures as traveling waves. Under which conditions these appear is an open question.

\section*{Author contributions}
\textcolor{black}{
All authors conceived the research, discussed the results and wrote the manuscript.
}

\section*{Conflicts of interest}
There are no conflicts to declare.
% \vspace{0.5cm}
\section*{Acknowledgements}
This work was supported by JSPS KAKENHI Grant Number JP17K05607 and ANR THEMA. HN acknowledges the visiting professorship program of University of Lyon 1.

%%%REFERENCES%%%
\bibliography{Binding_Thermalized_Activemembrane_curvature-inducing_Proteins} %You need to replace "rsc" on this line with the name of your .bib file
\bibliographystyle{apsrev4-1}
 %the RSC's .bst file

\end{document}